\pdfoutput=1
\documentclass[twocolumn,superscriptaddress,aps,preprintnumbers,amsmath,amssymb,prd,nofootinbib]{revtex4-2}

\usepackage{amsmath}
\usepackage{amssymb}
\usepackage{amsfonts}
\usepackage[pdftex]{graphicx}
\usepackage{xcolor}
\usepackage{xfrac}
\usepackage{comment}
\usepackage{pifont}
\usepackage{physics}
\usepackage{fourier}
\usepackage{hyperref}
\usepackage{bm}
\usepackage{enumitem}
\usepackage{cancel}
\usepackage[export]{adjustbox}

\definecolor{rossoferrari}{HTML}{D9073D}
\definecolor{mediumblue}{HTML}{0000CD}
\definecolor{forestgreen}{HTML}{228B22}
\definecolor{desy_blue}{HTML}{009EE2}
\definecolor{desy_orange}{HTML}{FD8800}
\definecolor{light_pink}{rgb}{1,0.4,0.4}
\definecolor{light_blue}{rgb}{0.284602,0.317763,0.963947}
\hypersetup{setpagesize=false,bookmarksnumbered=true,bookmarksopen=true,colorlinks=true,linkcolor=light_blue,urlcolor=rossoferrari,citecolor=rossoferrari,linktocpage=false}

\newcommand{\TeV}{\,\mathrm{TeV}}

\def\Mpl{M_\text{Pl}}

\newcommand{\lmk}{\left(}  
\newcommand{\rmk}{\right)}

\newcommand{\lhk}{\left \{}  
\newcommand{\rhk}{\right \}}

\newcommand{\g}{\, {\rm g}}

\def\beq#1\eeq{\begin{align}#1\end{align}}

\newcommand{\eq}[1]{Eq.~(\ref{#1})}

\begin{document}


\title{
Super-slow phase transition catalyzed by BHs and the birth of baby BHs
}

\author{Ryusuke Jinno}
\email{ryusuke.jinno@resceu.s.u-tokyo.ac.jp}
\affiliation{Research Center for the Early Universe (RESCEU), University of Tokyo, Bunkyo, Tokyo 113-0033, Japan}

\author{Jun'ya Kume}
\email{junya.kume@unipd.it}
\affiliation{Dipartimento di Fisica e Astronomia ``G. Galilei'', Universit\`a degli Studi di Padova, via Marzolo 8, I-35131 Padova, Italy}
\affiliation{INFN, Sezione di Padova, via Marzolo 8, I-35131 Padova, Italy}
\affiliation{Research Center for the Early Universe (RESCEU), University of Tokyo, Bunkyo, Tokyo 113-0033, Japan}

\author{Masaki Yamada}
\email{m.yamada@tohoku.ac.jp}
\affiliation{Department of Physics, Tohoku University, Sendai, Miyagi 980-8578, Japan}
\affiliation{FRIS, Tohoku University, Sendai, Miyagi 980-8578, Japan}

\preprint{TU-1209}
\preprint{RESCEU-18/23}

\date{\today}


\begin{abstract}
\noindent
We discuss the unique phenomenology of first-order phase transitions catalyzed by primordial black holes (BHs). If the number of BHs within one Hubble volume is smaller than unity at the time of bubble nucleation, each bubble catalyzed around them can expand to the Hubble size, and the universe is eventually filled with true vacuum much after nucleation. This super-slow transition predicts enhanced gravitational wave signals from bubble collisions and can be tested in future observations. Moreover, the remaining rare false vacuum patches give birth to baby BHs, which can account for the abundance of dark matter in our universe.
\end{abstract}


\maketitle


\section{Introduction}
Scalar fields are ubiquitous in physics beyond the Standard Model (SM) and may have a nontrivial vacuum structure.
It is expected that the Universe has experienced several phase transitions during its thermal history.
However, the vacuum nucleation rate can be comparable or smaller than the present age of the Universe.
For example, in the SM, the lifetime of the electroweak vacuum is slightly longer than the present age of the Universe~\cite{Sher:1988mj,Arnold:1989cb,Altarelli:1994rb,Espinosa:1995se,Casas:1996aq,Hambye:1996wb,Isidori:2001bm,Espinosa:2007qp,Ellis:2009tp,Bezrukov:2012sa,Bednyakov:2015sca,EliasMiro:2011aa,Degrassi:2012ry,Buttazzo:2013uya,Branchina:2013jra} (see, e.g., Ref.~\cite{Markkanen:2018pdo}).
In such a case, vacuum nucleation is expected to be triggered if the tunneling action is reduced by a factor of $\mathcal{O}(1\,\text{-}\, 10)$. 
This can occur around compact objects such as black holes (BHs).

Several studies have been conducted on the enhancement of vacuum nucleation rate around a BH~\cite{Hiscock:1987hn,Berezin:1987ea, Arnold:1989cq, Berezin:1990qs, Samuel:1992wt, Gomberoff:2003zh, Garriga:2004nm, Gregory:2013hja, Burda:2015isa, Burda:2015yfa,Burda:2016mou,  Chen:2017suz, Gorbunov:2017fhq, Canko:2017ebb, Mukaida:2017bgd, Kohri:2017ybt,Billam:2018pvp,Oshita:2019jan,Gregory:2020cvy,Hayashi:2020ocn,El-Menoufi:2020ron,Gregory:2020hia,Shkerin:2021zbf,Shkerin:2021rhy,Strumia:2022jil,Briaud:2022few,Gregory:2023kam}. 
Although it is under debate how to properly treat the thermal effect of Hawking radiation, 
it is still reasonable to expect that the vacuum nucleation rate around BHs is strongly enhanced by the strong gravitational field, at least for bubbles whose radii are comparable to those of the BHs.
Moreover, such an enhancement can be even stronger for compact objects without a horizon because the Bekenstein entropy is absent in these cases~\cite{Oshita:2018ptr}. 

\begin{figure*}[ht]
\centering
\includegraphics[width=1.1\columnwidth,valign=m]{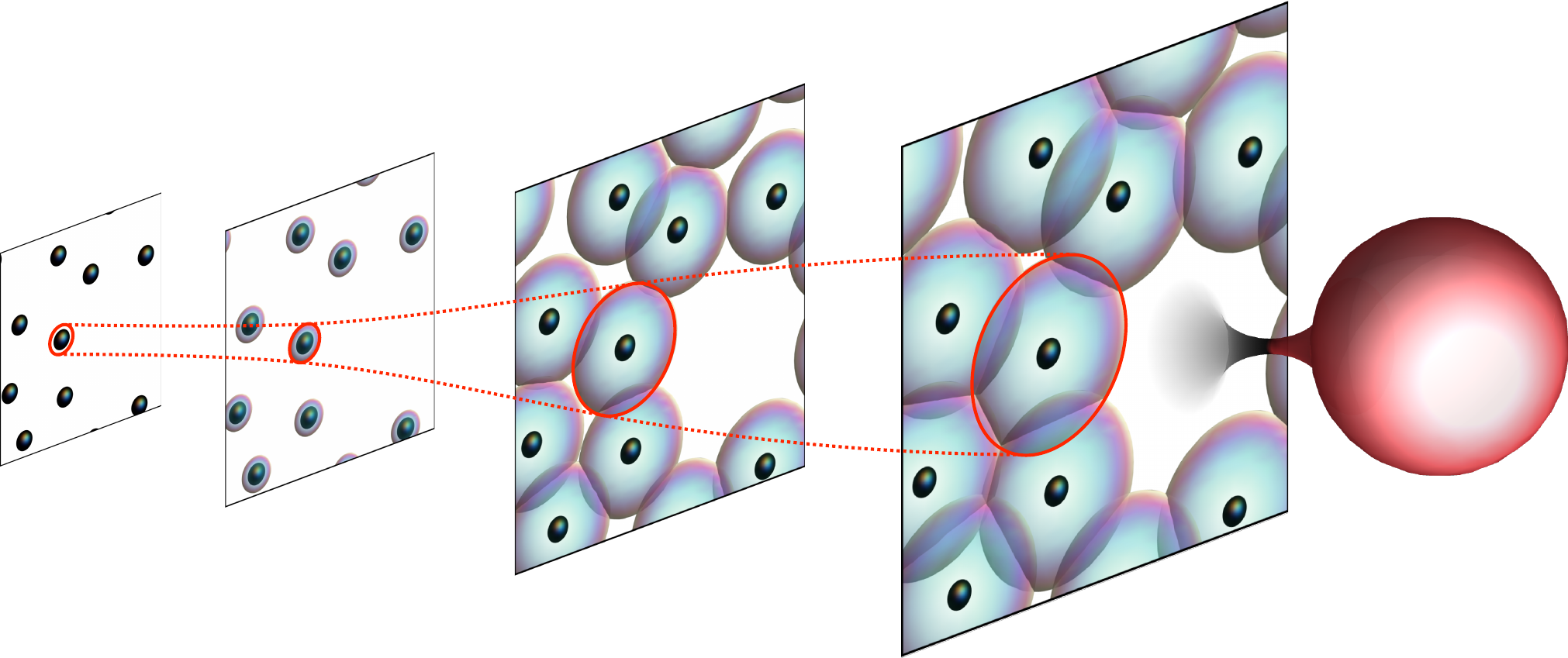}
\hskip 1cm
\includegraphics[width=0.8\columnwidth,valign=m]{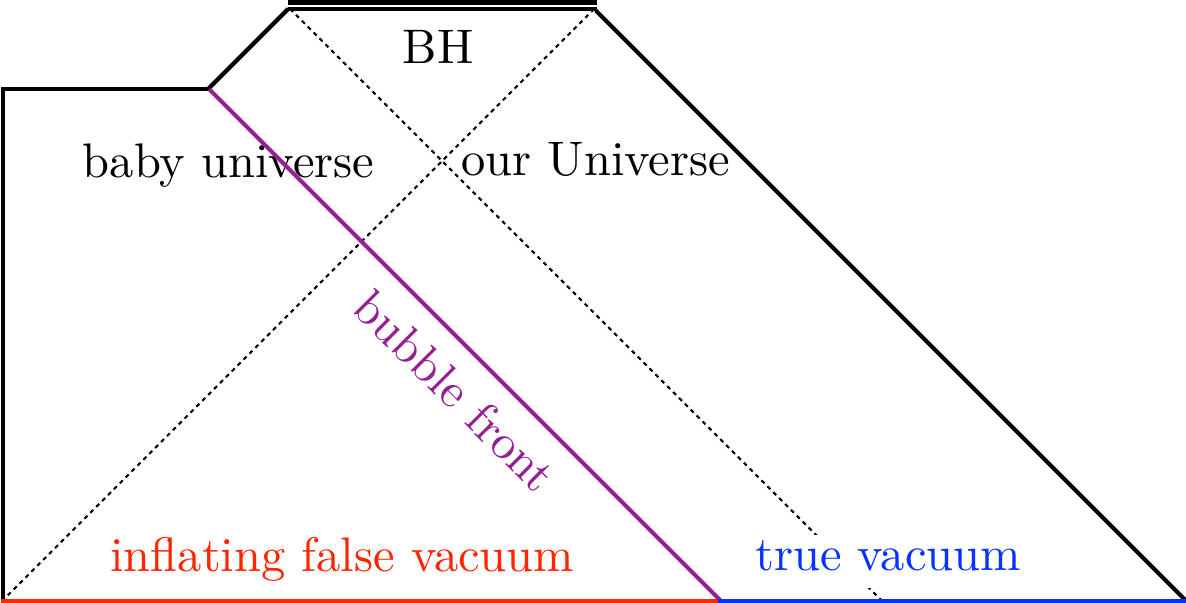}
\caption{
(Left) Schematic of cosmological evolution in our scenario. 
The black dots represent BHs, while the spheres surrounding them represent true vacuum regions.
The red circles on each sheet connected by dotted lines represent the Horizon scale at each cosmic time. 
The red sphere represents an inflating region that is connected to our universe by a wormhole.
(Right)
Penrose diagram starting from $t \sim H_{\rm eq}^{-1}$, when the vacuum energy dominates in the false vacuum region.
}
\label{fig:schematic}
\end{figure*}

In this letter, we consider a unique 
phenomenology of vacuum nucleation triggered around primordial black holes (PBHs), 
assuming that the false-true vacuum structure of the scalar potential is maintained in the course of evolution.
If the average number of PBHs within one Hubble volume is lower than unity at the bubble nucleation, bubbles can expand to the Hubble size.
The Hubble patches that contain BHs can transition to the true vacuum, whereas the remainder of the Hubble patches remain in the false vacuum (left panel of Fig.~\ref{fig:schematic}).
The phase transition is eventually completed only after the average number of BHs within one Hubble volume is of the order of unity.
This super-slow phase transition scenario provides a unique signal in the gravitational waves (GWs) from bubble collisions~\cite{Kosowsky:1991ua,Kosowsky:1992rz,Kosowsky:1992vn,Kamionkowski:1993fg,Huber:2008hg,Jinno:2016vai,Jinno:2017fby}.

Moreover, even though the phase transition is successfully completed,
some rare regions remain in the false vacuum and inflate forever (the red region in the left panel of Fig.~\ref{fig:schematic}).
These regions result in baby universes that are connected to our (parent) universe by wormholes and are eventually seen as BHs by an outside observer~\cite{Kodama:1981gu,Berezin:1982ur,Blau:1986cw,Maeda:2009tk}. (See also Refs.~\cite{Garriga:2015fdk,Deng:2016vzb,Deng:2017uwc,Deng:2018cxb} for wormhole formation from collapsing domain walls or vacuum bubbles and Refs.~\cite{Sato:1981bf,Maeda:1981gw,Sato:1981gv,Kodama:1981gu,Kodama:1982sf} for that after a first-order phase transition (FOPT) in different contexts.) 
We describe the causal structure of this phenomenon using the Penrose diagram shown in the right panel of Fig.~\ref{fig:schematic}, assuming spherical symmetry around the center of the inflating region for simplicity.
The purple curve represents the vacuum bubble front that sweeps toward the center of the coordinate system (corresponding to the left most part), while our universe is in the true vacuum region.
A baby universe is created because of inflation in the false vacuum region and eventually becomes disconnected from our universe by the event horizons (dotted lines) of the baby BH.

This provides a unique scenario for the generation of large baby PBHs from the phase transition triggered by the original PBHs. 
Similar but different scenarios for PBH formation during a FOPT have been previously considered~\cite{Hawking:1982ga,Moss:1994iq,Khlopov:1998nm,Jedamzik:1999am,Lewicki:2019gmv,Gross:2021qgx,Baker:2021nyl,Kawana:2021tde,Liu:2021svg,Baker:2021sno,Jung:2021mku,Hashino:2021qoq,Huang:2022him,He:2022amv,Hashino:2022tcs,Kawana:2022olo,Lewicki:2023ioy,Gouttenoire:2023naa,Salvio:2023ynn,Baldes:2023rqv}. 
In these studies, the authors considered the collapse of overdense regions following the collision of the bubbles.
Their mechanism may depend on the details of the phenomenology of the bubbles and surrounding plasma as well as the assumption of spherical symmetry, whereas in our case, BHs appear in regions where bubble collision is not complete.
In the mechanism we propose, inflation naturally occurs in the false vacuum regions larger than the Hubble horizon, even if they are not spherically symmetric.
Furthermore, our argument does not rely on the complicated and nontrivial dynamics of the collapse of overdense regions, and thus we believe the present letter offers a rather robust discussion on PBH formation.

\section{Super-slow phase transition catalyzed by PBHs}

In this letter, we consider PBHs as the catalysts of a first-order phase transition.
As mentioned above, the nucleation rate can be drastically enhanced around the BH if the radius of BH is comparable to the bubble radius for vacuum nucleation. 
For a sector undergoing the catalyzed transition, we assume a scalar field satisfying the following conditions: 
i)The scalar potential develops a meta-stable vacuum and true vacuum before the time of catalyzed nucleation, and our Universe is trapped in the former.
This vacuum structure is maintained until the transition is completed.
ii)The vacuum nucleation rate in a homogeneous background is too small for the transition to occur, but it is sufficiently enhanced around the PBHs so that bubbles can nucleate.
iii)The cosmological constant is vanishingly small in the true vacuum.
Note that i) is a rather nontrivial assumption,
especially for the case of a thermal transition.
As a possible and simple realization, we may envision a ``dark'' scalar with the zero-temperature potential possessing a false-true vacuum structure, so that its potential is (almost) unaffected by the thermal history of the Universe.%
\footnote{
One can also consider a hot dark sector, where the false vacuum is stabilized by thermal effect at a high temperature. The phase transition can happen around PBHs once the temperature decreases enough to make the false vacuum metastable. 
}

Under these assumptions, we arrive at a situation where the true vacuum bubbles nucleate only around the PBHs.
For simplicity, we take the initial mass of PBHs, $M_{\rm PBH,i}$, to be monochromatic so that the bubbles nucleate simultaneously around all PBHs. 
We denote the time of PBH formation and bubble nucleation respectively as $t_*$ and $t_{\rm nuc}$ ($\ge t_*$), and the PBH number density as $n(t)$ ($\propto a^{-3}(t)$).
Let us define $\epsilon$ as the average number of PBHs within one Hubble volume at $t = t_{\rm nuc}$: 
\beq
 \epsilon \equiv \frac{4 \pi n(t_{\rm nuc})}{3 H^3(t_{\rm nuc})}.
\eeq
The case $\epsilon < 1$ is the focus of this study.
In this case, the true vacuum bubbles do not fill the universe within one Hubble time after $t = t_{\rm nuc}$.
The bubbles expand with velocity $v_b$ ($\sim 1$), and their radii become of the order of the Hubble length.
Subsequently, these Hubble-sized bubbles collide with each other, and the true vacuum slowly fills the universe.

Note that $\epsilon$ is related to the energy fraction of the universe $\beta_{\rm PBH}$, which collapses into a PBH at the temperature $T_*$, as described below: 
\begin{align}
\epsilon = \frac{\beta_{\rm PBH}}{\gamma}\lmk\frac{T_*}{T_{\rm nuc}}\rmk^3,
\label{eq:betaPBH}
\end{align}
where $\gamma$ is the ratio of the PBH mass to the horizon mass.
Here, for simplicity, we assume that the initial mass of the PBH is $M_{\rm PBH,i} \simeq 4\pi \gamma \Mpl^2 / H_*$, which holds for PBH formation from the collapse of overdense regions in a radiation-dominated epoch. 
Hereafter, we consider the fiducial value of $\gamma$ as 0.2~\cite{Carr:1975qj}.
Our scenario can be applied to other PBH formation mechanisms with some trivial corrections. 

We quantitatively discuss our scenario according to the percolation theory.
The probability of finding a spatial point in false vacuum at time $t$ is given by $P(t) = e^{-I(t)}$, where 
\beq
 &I(t) = \frac{4\pi}{3} \int_0^t dt' \Gamma (t') 
 a^3 (t') r_{\rm bubble}^3 (t, t'), 
 \\
 &r_{\rm bubble} (t,t') = v_b \int_{t'}^t \frac{dt''}{a(t'')}. 
\eeq
Here, $\Gamma(t)$ denotes the bubble nucleation rate per unit time and volume.
We assume $v_b = {\rm const.}$ for simplicity.
In our case, bubble nucleation occurs at $t = t_{\rm nuc}$ around the BHs such that 
\beq
\Gamma(t') = \delta(t' - t_{\rm nuc}) n (t_{\rm nuc}).
\label{eq:Gamma}
\eeq
Using the above expressions, the time $t_{\rm col}$ of the bubble collision and thus the completion of the phase transition can be defined as $I(t_{\rm col}) = 1$.

If the energy density of the false vacuum region is dominated by the false vacuum energy by the time $t = t_{\rm col}$, that region experiences eternal inflation, and the entire universe is never filled with true vacuum bubbles.
Instead, we assume that the false vacuum region is dominated by the radiation energy until $t = t_{\rm col}$.  
This condition can be represented as follows: 
\beq
\alpha_{\rm col} \equiv \left. \frac{\rho_{\rm vac}}{\rho_{\rm rad}} \right\vert_{t = t_{\rm col}} \ll 1, 
\eeq
where $\rho_{\rm vac}$ is the false vacuum energy in the transition sector and $\rho_{\rm rad}$ is the radiation energy in the SM sector. 
In this case, the true vacuum bubbles eventually collide with other bubbles, and the phase transition can be completed within a finite time.

Under the assumption of radiation domination, $I(t)$ is evaluated as
\begin{align}
I(t) = \epsilon v_b^3\lhk\lmk\frac{t}{t_{\rm nuc}}\rmk^{1/2} - 1\rhk^3.
\end{align}
Thus, we obtain 
\beq
t_{\rm col} = \lmk 1 +  \frac{1}{\epsilon^{1/3} v_b}\rmk^{2}t_{\rm nuc}.
\label{eq:col-tr}
\eeq

The growth rate of the true vacuum fraction can be quantified as
\begin{align}
\beta_P(t) \equiv - \frac{d \ln P}{dt} = \dot{I} = \frac{3I(t)v_b}{a(t)r_{\rm bubble}(t, t_{\rm nuc})}. 
\end{align}
Because $a(t)r_{\rm bubble}(t, t_{\rm nuc}) = 2v_b t (1 - (t_{\rm nuc}/t)^{1/2})$, we obtain $\beta_P / H \simeq 3 I(t) \propto t^{3/2}$ during the radiation-dominated era. 
In particular, we obtain 
\begin{align}
\beta_P(t_{\rm col}) = 3H(t_{\rm col})\lhk1 - \lmk 1 +  \frac{1}{\epsilon^{1/3} v_b}\rmk^{-1}\rhk^{-1},
\label{eq:betacol}
\end{align}
which can be regarded as the inverse duration of the phase transition in this scenario.
One sees that $\epsilon \to 0$ gives $\beta_P (t_{\rm col}) / H (t_{\rm col}) = 3$, and also that $\epsilon = 0.1$ already gives $\beta_P (t_{\rm col}) / H (t_{\rm col}) \simeq 4.4$, only $\simeq 50 \%$ different from the $\epsilon \to 0$ limit.
Our following discussion will be valid as long as such a small value of $\beta_P (t_{\rm col}) / H (t_{\rm col})$ is realized.

Notably, $\beta_P/H_{\rm col} \simeq 3$ ($\epsilon \ll 1$) is actually the lowest allowed value for $\beta_P$ because the volume of the false vacuum region continues to decrease only if 
\beq
\frac{d}{dt} \lmk a^3(t) P(t) \rmk < 0 
\quad \leftrightarrow \quad
\frac{\beta_P(t)}{H(t)} > 3. 
\eeq
The feature $\beta_P/H_{\rm col} \simeq 3$ quantitatively represents the slowness of the phase transition.
Such a small value of $\beta_P/H_{\rm col}$ is difficult to realize in standard scenarios without fine-tuning but is sometimes motivated from a phenomenological point of view (see, e.g.,~\cite{Megevand:2016lpr,Kobakhidze:2017mru,Cai:2017tmh,Megevand:2017vtb,Ellis:2018mja,Wang:2020jrd,Athron:2022mmm,Athron:2023mer}).

\section{Birth of baby BHs}

Despite the successful completion of the phase transition, some rare regions could remain in a false vacuum and inflate forever.
These regions are seen as BHs by outside observers~\cite{Garriga:2015fdk,Deng:2017uwc,Deng:2020mds}.
Thus, we arrive at the scenario of ``the birth of baby BHs from vacuum decay catalyzed by parent BHs".

The number density of baby PBHs is estimated as follows: 
A false vacuum region can inflate if it is as large as the Hubble horizon for the false vacuum $H_{\rm fv}^{-1}$ at $t \gtrsim t_{\rm eq}$, where $t_{\rm eq}$ is the time at which the vacuum energy begins to dominate, and is defined by $\rho_{\rm rad}(t_{\rm eq}) = \rho_{\rm vac}$. 
The average number of parent PBHs within a volume of $(4\pi/3) H_{\rm fv}^{-3}$ is given by 
\beq
N_{\rm PBH} (t_{\rm eq}) \simeq 
\epsilon \lmk \frac{a_{\rm eq}}{a_{\rm nuc}} \rmk^3 
\simeq v_b^{-3} \alpha_{\rm col}^{-3/4},
\eeq
meaning that the value of $\epsilon$ can be traded for $\alpha_{\rm col}$ and $t_{\rm nuc}$, the latter being implicitly in $a_{\rm nuc}$.
Assuming that the number of PBHs within the region follows a Poisson distribution, we estimate the probability that no true vacuum bubble forms within the same volume as $P \simeq e^{-N_{\rm PBH} (t_{\rm eq})}$. 
The number density of the baby PBHs is then given by $P/((4\pi/3) H_{\rm fv}^{-3})$.
The inflating false vacuum region can eventually be perceived as a BH by distant observers.
Its mass is estimated as 
\beq
M_{\rm baby} \simeq \frac{4\pi}{3} H_{\rm fv}^{-3} \cdot 3 \Mpl^2 H_{\rm fv}^2
\simeq 2.8 \times 10^{-7} M_{\odot}\lmk\frac{\rm TeV^4}{\rho_{\rm vac}}\rmk^{1/2}.
\label{eq:Mbaby}
\eeq

The abundance of baby BHs is also subject to observational constraints that depend on their masses. 
In particular, for $10^{-15}M_{\odot} \lesssim M_{\rm baby} \lesssim 10^{-10}M_{\odot}$, baby BHs are allowed to constitute the entire dark matter (DM) of our universe.
For comparison with the strongest observational bounds for a monochromatic PBH mass, as shown in Fig.~18 of Ref.~\cite{Carr:2020gox}, it is convenient to introduce a parameter $\beta_{\rm bBH}$, which characterizes the fraction of energy in the universe occupied by baby BHs during their formation
\begin{align}
\beta_{\rm bBH} \equiv \frac{\rho_{\rm bBH}}{3M_{\rm Pl}^2H_{\rm fv}^2} = P = \exp\lmk - v_b^{-3}\alpha_{\rm col}^{-3/4}\rmk.
\label{eq:beta_bBH}
\end{align}
where $M_{\rm Pl} \equiv (8 \pi G)^{-1/2}$ is the reduced Planck mass with $G$ being the Newtonian constant.
For example, when $\rho_{\rm vac} \sim (10^5{\rm TeV})^4$, the baby BH mass is $M_{\rm baby} \sim 10^{10}$g, and the constraint from Big-Bang nucleosynthesis (BBN) $\beta_{\rm bBH} \lesssim 10^{-24}$ yields $v_b^4\alpha_{\rm col} \lesssim 5 \times 10^{-3}$. 
Constraints on PBH abundance typically result in $\alpha_{\rm col} < \mathcal{O}(10^{-2})$.
Detailed values of the upper bound will be provided shortly (see Fig.~\ref{fig:alpha_bound}). 

Note that the largest allowed value of $\alpha_{\rm col}$ can be increased by lowering the wall velocity $v_b$, which may be achieved by the friction between the wall and fluid. 
In most realistic cases, however, we expect the velocity to eventually reach $v_b \sim 1$ as any radiation coupled to the transition sector must be extremely dilute from the beginning in our setup so as not to spoil the metastable vacuum structure.
Therefore, we use $v_b = 1$ as the fiducial value.

\section{GW signals}

A stochastic gravitational wave background (SGWB) is produced by complicated dynamics during the cosmological phase transition, which is typically classified into contributions from bubble collisions, sound waves, and turbulence.
In our scenario, in which the Hubble-sized bubbles collide over the Hubble time, the dominant source of SGWB is expected to be the collision of the bubble walls~\cite{Kosowsky:1991ua,Kosowsky:1992rz,Kosowsky:1992vn,Kamionkowski:1993fg,Huber:2008hg,Jinno:2016vai,Jinno:2017fby}.\footnote{Depending on the model, the fluid contribution could be comparable to Eq.~\eqref{eq:GWamp}.
In other words, our estimate of SGWB is relatively conservative because it is based solely on bubble collisions.
}
The amplitude and characteristic frequency of the SGWB generated by bubble collisions can be roughly estimated as~\cite{Caprini:2015zlo}
\begin{align}
\Omega_{\rm gw} h^2 &\sim 1.6 \times 10^{-5}\lmk\frac{100}{g_*(T_{\rm col})}\rmk^{1/3}\lmk\frac{H_{\rm col}}{\beta_P}\rmk^2 \lmk\frac{\kappa_{\phi}\alpha_{\rm col}}{1 + \alpha_{\rm col}}\rmk^2,
\label{eq:GWamp}\\
f &\sim 1.6 \times 10^{-3}{\rm Hz}\lmk \frac{g_*(T_{\rm col})}{100}\rmk^{1/6}\frac{\beta_P}{H_{\rm col}}\frac{T_{\rm col}}{10\TeV}
\label{eq:GWfreq},
\end{align}
where $\kappa_{\phi}$ represents the fraction of energy transferred from vacuum energy to the scalar kinetics and gradient.
In contrast to the standard scenario of cosmological FOPT, the super-slow phase transition scenario predicts $\beta_P/H_{\rm col} \simeq 3 = \mathcal{O}(1)$. 
Therefore, the amplitude of the SGWB from bubble collision can be within the reach of future observation of the Pulsar Timing Arrays (PTAs)~\cite{Manchester:2012za, McLaughlin:2013ira, Kramer:2013kea, Manchester:2013ndt, Janssen:2014dka} and planned detectors, such as the Einstein Telescope~\cite{Punturo:2010zz, Hild:2010id}, Cosmic Explorer~\cite{LIGOScientific:2016wof, Reitze:2019iox}, LISA~\cite{https://doi.org/10.48550/arxiv.1702.00786, Baker:2019nia}, and DECIGO~\cite{Seto:2001qf, Kawamura:2006up}. 

Based on \eq{eq:col-tr}, the temperature at the bubble collision $T_{\rm col}$ can be expressed via the PBH parameters by using \eq{eq:betaPBH} and representing $T_*$ in terms of $M_{\rm PBH,i}$ as 
\begin{align}
T_{\rm col} &\simeq \gamma^{1/6}v_b\beta_{\rm PBH}^{1/3}\sqrt{\frac{4\pi M_{\rm Pl}^3}{M_{\rm PBH,i}}}\lmk\frac{90}{\pi^2g_*(T_*)}\rmk^{1/4}\notag\\
&\simeq 23 \TeV \, 
\lmk\frac{g_*(T_*)}{100}\rmk^{-1/4}
\lmk\frac{\beta_{\rm PBH}}{10^{-21}}\rmk^{1/3}\lmk\frac{M_{\rm PBH,i}}{10^9 \g}\rmk^{-1/2},
\label{eq:T_col}
\end{align}
for $\epsilon \ll 1$.
Using this expression, the peak frequency Eq.~\eqref{eq:GWfreq} can be expressed in terms of the PBH parameters, whereas the strength of the phase transition $\alpha_{\rm col} = \rho_{\rm vac}/\rho_{\rm rad}(T_{\rm col})$ depends on the details of the scalar sector and requires an additional assumption, which we will specify shortly.

\subsection{Predictions and existing constraints}

During the bubble nucleation catalyzed by the PBHs, the radius of the vacuum bubbles should be comparable to the BH radius $M_{\rm PBH}/M_{\rm Pl}^2$.
Although the relationship between the radius of the vacuum bubble and vacuum energy $\rho_{\rm vac}$ depends on the model details, we simply assume the following relationship: 
\beq
\rho_{\rm vac}^{1/4} \sim M_{\rm Pl}^2/M_{\rm PBH},
\label{eq:radius}
\eeq
which should be sufficient for an order-of-magnitude estimate.
From \eq{eq:Mbaby}, the mass of the baby BHs is expressed as $M_{\rm baby} \simeq 5.0\times10^{24}{\rm g} \lmk M_{\rm PBH}/(10^9\,{\rm g} ) \rmk^{2}$.

If we further assume $T_{\rm nuc} \gg T_{\rm ev}$, where $T_{\rm ev}$ is the evaporation temperature of the PBH, $M_{\rm PBH}$ is approximately equal to $M_{\rm PBH,i}$. 
Therefore, once $(M_{\rm PBH, i}, \beta_{\rm PBH})$ are given, $T_{\rm col}$ and $\alpha_{\rm col}$ are fixed such that
\beq
\alpha_{\rm col} 
&\sim \frac{1}{3}\lmk\frac{g_*(T_*)}{g_*(T_{\rm col})}\rmk v_b^{-4}\beta_{\rm PBH}^{-4/3}\frac{1}{(4\pi\gamma^{1/3})^2}\lmk\frac{M_{\rm Pl}}{M_{\rm PBH, i}}\rmk^2\notag\\
&\sim 10^{-3}
\lmk\frac{g_*(T_*)}{g_*(T_{\rm col})}\rmk 
\lmk\frac{\beta_{\rm PBH}}{10^{-21}}\rmk^{-4/3}\lmk\frac{M_{\rm PBH,i}}{10^9 \g}\rmk^{-2}.
\label{eq:alpha_col}
\eeq

Now we consider the upper bound on $\alpha_{\rm col}$, which originates from the constraints imposed on the abundance of baby BHs. 
The mass of the baby BHs is given by 
\begin{align}
M_{\rm baby} \simeq 5.0\times10^{24}{\rm g} 
\lmk\frac{M_{\rm PBH,i}}{10^9{\rm g}}\rmk^{2},
\label{eq:bBHmass}
\end{align}
under the assumptions described by Eq.~\eqref{eq:radius} and $T_{\rm nuc} \gg T_{\rm ev}$. 
Using $\beta_{\rm bBH}$ defined in \eq{eq:beta_bBH} for the abundance of baby BHs, the constraints as provided in Ref.~\cite{Carr:2020gox} yields an upper bound on $\alpha_{\rm col}$ for a given $M_{\rm PBH,i}$. 

\begin{figure}
\centering
\includegraphics[width=1\columnwidth]{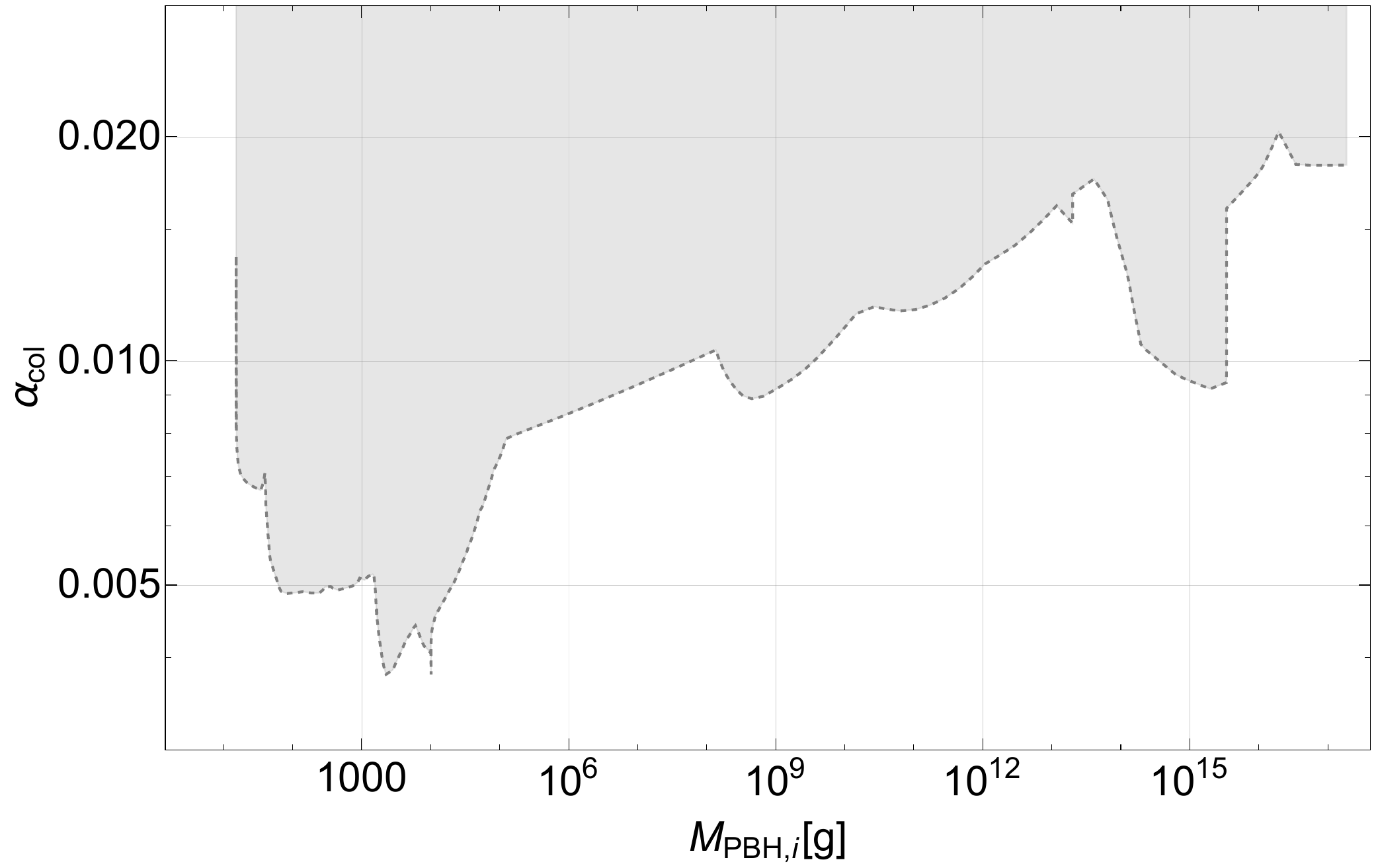}
\hspace{0.3cm}
\includegraphics[width=1\columnwidth]{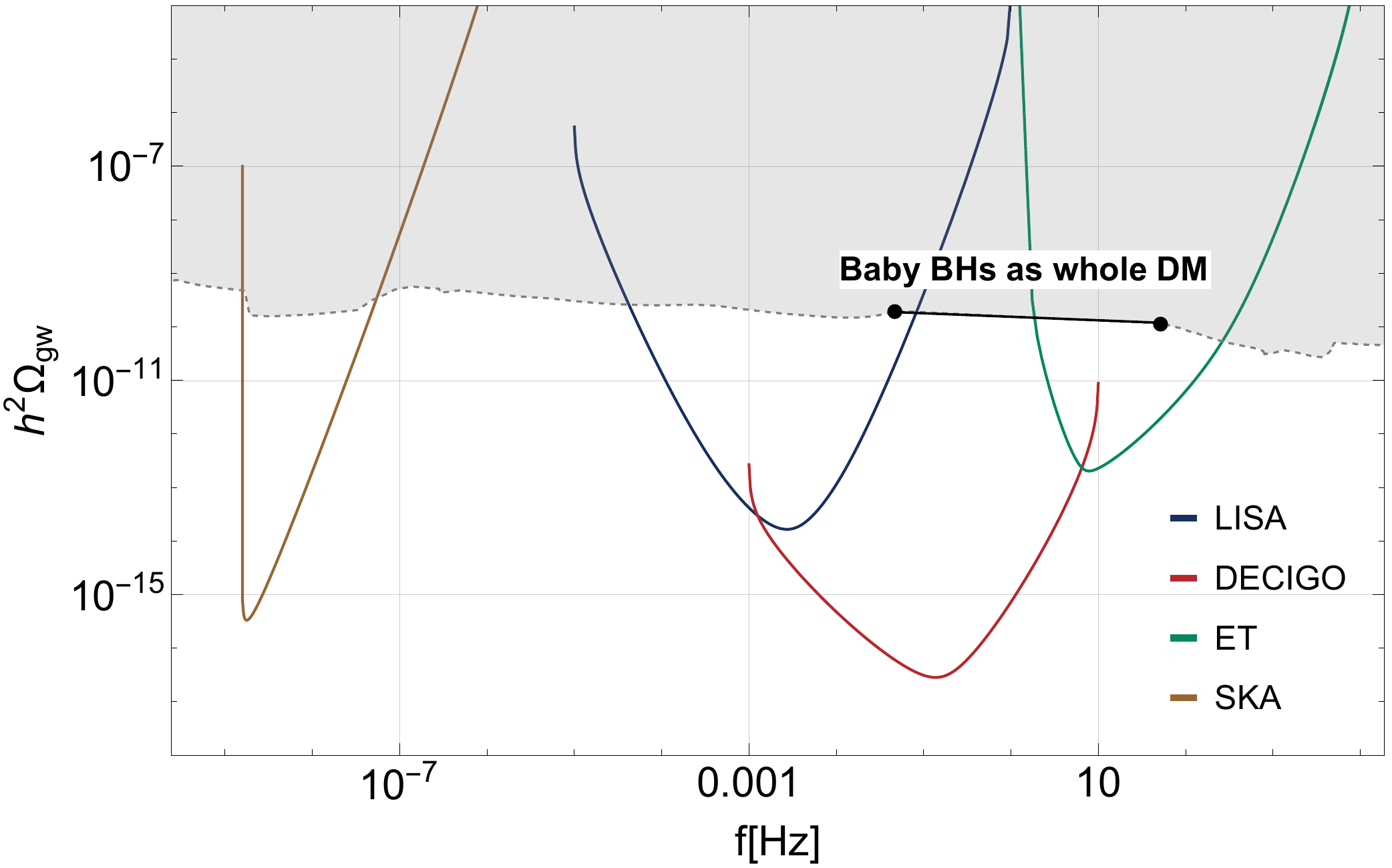}
\caption{({\it Top panel})
Upper bound on $\alpha_{\rm col}$ derived from the observational constraints imposed on $\beta_{\rm bBH}$ based on Eqs.~\eqref{eq:beta_bBH} and ~\eqref{eq:bBHmass}.
({\it Bottom panel}) Projection of the maximally allowed value of $\alpha_{\rm col}$ for each $M_{\rm PBH,i}$ onto the $(f, \Omega_{\rm gw})$ plane.
For both plots, the upper shaded region is excluded in our scenario.
}
\label{fig:alpha_bound}
\end{figure}

The excluded region (derived from $v_b = 1$) is shown in the top panel of Fig.~\ref{fig:alpha_bound} over the mass range relevant to the SGWB observations.
If the mass of the baby PBHs is less than about $10^{-15}M_{\odot}$, they evaporate by the present epoch.
In this case, BBN~\cite{Carr:2009jm}, anisotropies of cosmic microwave background (CMB)~\cite{Acharya:2020jbv}, the galactic and extragalactic $\gamma$-ray (GGB~\cite{Carr:2016hva} and EGB~\cite{Carr:2009jm}), and cosmic ray backgrounds~\cite{Boudaud:2018hqb} attain the tightest bounds on their abundance at formation.
However, baby BHs survive to the present if they are heavier than $10^{-18}M_{\odot}$.
The tightest (and relatively secure) bound on their abundance comes from gravitational lensing~\cite{Niikura:2017zjd,Smyth:2019whb, Niikura:2019kqi, EROS-2:2006ryy, Oguri:2017ock}, dynamical friction and its influence on large-scale structures~\cite{Carr:2018rid}, accretion~\cite{Inoue:2017csr,Serpico:2020ehh}, and gravitational waves~\cite{LIGOScientific:2019kan}.
In particular, for $10^{-15}M_{\odot} \lesssim M_{\rm baby} \lesssim 10^{-10}M_{\odot}$, baby BHs are allowed to constitute the entire DM of our universe.
The maximally allowed value of $\alpha_{\rm col}$ is approximately $\mathcal{O}(10^{-2})$ for all the masses described here.

Under our assumptions (Eq.~\eqref{eq:radius} and $T_{\rm nuc} \gg T_{\rm ev}$), the peak frequency and amplitude of SGWB can be determined using Eqs.~\eqref{eq:GWamp} and \eqref{eq:GWfreq}.
With the parameters fixed at $\gamma = 0.2$, $g_*(T_*) = g_*(T_{\rm col}) = 100$ and $\beta/H_p = 3$, the largest amplitudes of $\Omega_{\rm gw}$ in our scenario are derived from the upper bound of $\alpha_{\rm col}$ for a given $M_{\rm PBH,i}$. 
(To be precise, $g_*(T_{\rm col}) \sim 10$ is expected for the PTA band, although this difference is only slight.)
As shown in the bottom panel of Fig.~\ref{fig:alpha_bound}, the amplitudes are compared with the power-law integrated sensitivity curves of the future SGWB observations. 
As the window of the baby-BH DM is described by $10^5{\rm g} \lesssim M_{\rm PBH, i} \lesssim 10^8{\rm g}$,
a particular scenario exists, where the baby PBHs can explain the abundance of DM in the universe while the SGWB from the bubble collisions can be observed in the laser interferometer band. 

\begin{figure}
\centering
\includegraphics[width=1\columnwidth]{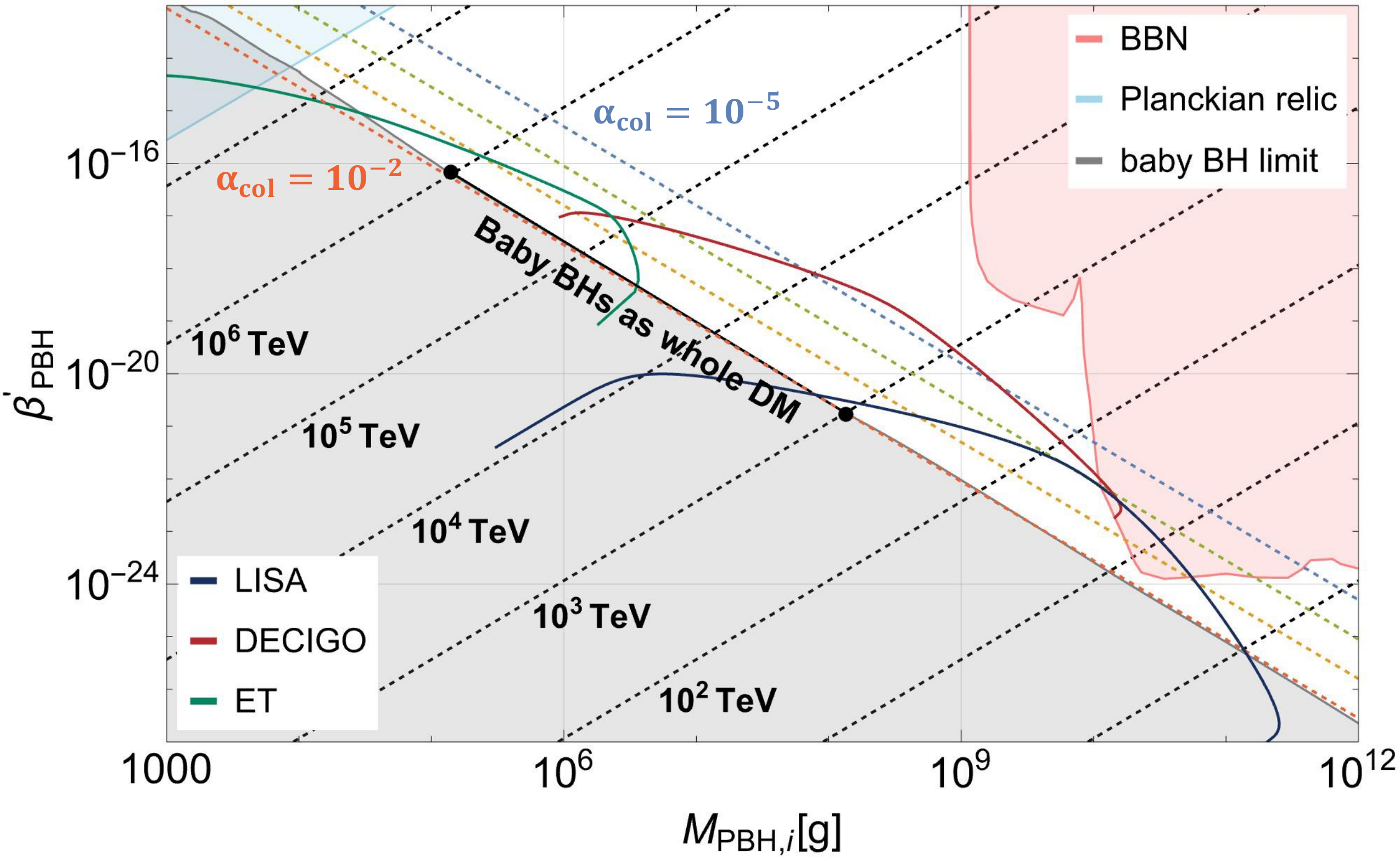}
\vspace{0.3cm}
\\
\includegraphics[width=1\columnwidth]{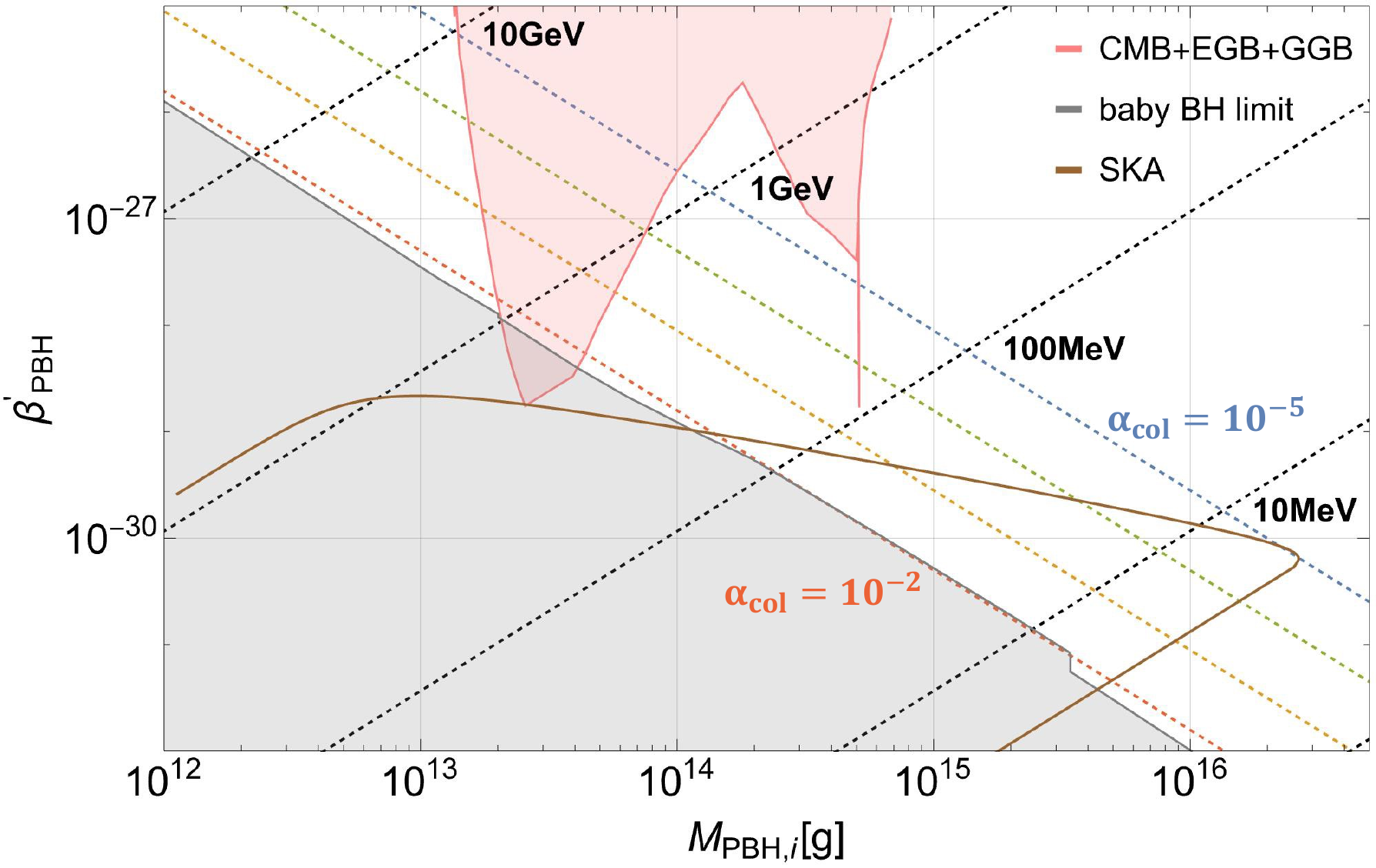}
\caption{Contours of $\alpha_{\rm col}$ (color-dashed) and $T_{\rm col}$ (black-dashed) drawn in the PBH parameter space $(M_{\rm PBH,i}, \beta_{\rm PBH}^{\prime})$.
Red and light-blue shaded regions are the constraints on the (parent) PBHs from BBN~\cite{Carr:2009jm} and Planckian relics~\cite{Carr:1994ar}, respectively. 
The left-bottom gray-shaded region is excluded in our scenario by the constraints on baby BHs. 
The solid-colored lines are the projection of the power-law integrated sensitivity curves of the future GW observations. 
The bottom panel shows the results in the PTA band, where $g_*(T_{\rm col})/g_*(T_*) = 1/10$ and the $\alpha_{\rm col}$ contours are slightly shifted above.
}
\label{fig:T*_Ttr_LI}
\end{figure}

In Fig.~\ref{fig:T*_Ttr_LI}, we summarize how the (seed) PBH parameters $M_{\rm PBH, i}$ and $\beta_{\rm PBH}^{\prime} \equiv \gamma^{1/2}\beta_{\rm PBH}$ in our scenario can be constrained by future SGWB observations in combination with the other observational constraints discussed in Ref~\cite{Carr:2020gox}.
To show the accessible parameter region, we first plot the contours of $T_{\rm col}$ (black-dashed lines with $T_{\rm col} = 10^0, 10^1, \cdots, 10^7$\,TeV from bottom-right to top-left) and $\alpha_{\rm col}$ (color-dashed lines with $\alpha_{\rm col} = 10^{-2}, 10^{-3}, 10^{-4}$ and $10^{-5}$ from bottom-left to top-right) using Eqs.~\eqref{eq:T_col} and \eqref{eq:alpha_col}, respectively.
Subsequently, using Eq.~\eqref{eq:GWamp}, where $\beta_P/H_{\rm col} = 3$, $\kappa_{\phi} = 1$ and $g_*(T_{\rm col}) = 100$ (10) for the top panel (bottom panel) of Fig.~\ref{fig:T*_Ttr_LI}, the power-law integrated sensitivity curves of the detectors (acquired from Ref.~\cite{Schmitz:2020syl}) are projected onto the $(M_{\rm PBH, i}, \beta_{\rm PBH}^{\prime})$ plane.
That is, the lower-left side of these projected curves is the region that can be probed by the SGWB observations.

The gray (red and light-blue) shaded region is excluded because of the overproduction of baby (parent) BHs.
The black line connected by the black dots represents the region in which the DM can be explained by the baby BHs. 
This is within the sensitivity curves of the planned laser interferometers.

Finally, we should check if the condition $\epsilon < 1$ is consistently satisfied in the parameter space of our interest. 
The condition is translated into an upper bound on $\beta_{\rm PBH}'$ by using \eq{eq:betaPBH} as 
\beq
\beta_{\rm PBH}' < \gamma^{3/2}  
\lmk \frac{T_{\rm nuc}}{T_*} \rmk^{3} .
\eeq
As long as the transition temperature $T_{\rm nuc}$ is not significantly lower than the PBH formation temperature $T_*$, this condition can be met in the entire region plotted in Fig.~\ref{fig:T*_Ttr_LI}.
\footnote{
If nucleation occurs around the PBH evaporation time, $t_{\rm nuc} = t_{\rm evap}$, the condition $\epsilon < 1$ is violated in part of the parameter space.
In this case one needs a separate consideration.
}

\section{Discussion}

We have discussed the phenomenology of a first-order phase transition triggered around PBHs with a small number density.
This provides a novel scenario of a super-slow phase transition in which the bubbles grow to the Hubble size and collide over a time significantly longer than the Hubble time at nucleation.
We expect two distinct features in relation to the standard scenario of FOPT.
The first is the approximate delta-function feature of the effective nucleation rate \eq{eq:Gamma}.
The GW spectrum in this case is expected to be more sharply peaked than that with the ordinary exponential nucleation rate, reflecting the uniform bubble-size distribution~\cite{Jinno:2017ixd}.
Second, the duration of the phase transition is longer than one Hubble time for a small number density of parent BHs, which implies that the expansion of the universe affects the spectrum of GWs. 
From these results, we expect that our scenario is distinguishable from the standard scenario with a precise determination of the GW spectrum.
We leave the precise GW spectrum for future research. 

Throughout this letter, we specifically consider PBHs~\cite{Hawking:1971ei, Carr:1974nx, Carr:1975qj, Ivanov:1994pa, GarciaBellido:1996qt, Kawasaki:1997ju, Yokoyama:1998pt, Garriga:2015fdk, Garcia-Bellido:2017mdw, Deng:2017uwc, Hertzberg:2017dkh} as catalysts; however, other compact objects including monopoles~\cite{Vilenkin:2000jqa}, Q-balls~\cite{Coleman:1985ki, Kusenko:1997si,Enqvist:1997si,Enqvist:1998en,Kasuya:1999wu,Kasuya:2000wx,Kasuya:2001hg, Lee:1988ag, Kasuya:2015uka, Hong:2015wga}, oscillons~\cite{Bogolyubsky:1976nx,Segur:1987mg,Gleiser:1993pt,Copeland:1995fq,Gleiser:1999tj, Kasuya:2002zs,Gleiser:2004an,Fodor:2006zs,Hindmarsh:2006ur, Amin:2011hj, Mukaida:2016hwd}, boson stars (including axion stars)~\cite{Ruffini:1969qy, Hogan:1988mp,Kolb:1993zz,Seidel:1993zk,Hu:2000ke,Guzman:2006yc,Sikivie:2009qn,Liebling:2012fv, Guth:2014hsa, Eby:2014fya, Braaten:2015eeu, Braaten:2016kzc, Hui:2016ltb, Eby:2016cnq, Eby:2017teq, Eby:2018ufi}, gravastars~\cite{Mazur:2001fv,Mazur:2004fk} (see also Ref.~\cite{Carballo-Rubio:2017tlh}), neutron stars, and BH remnants~\cite{Aharonov:1987tp,Adler:2001vs} should function similarly if their gravitational potential is sufficiently strong~\cite{Samuel:1992wt,Oshita:2018ptr,Balkin:2021zfd}.\footnote{
See also Refs.~\cite{Kusenko:1997hj, Metaxas:2000qf, Pearce:2012jp} for Q-balls without gravity and Refs.~\cite{Steinhardt:1981ec, Steinhardt:1981mm, Jensen:1982jv,
Witten:1984rs, Yajnik:1986tg, Yajnik:1986wq, Dasgupta:1997kn,
Kumar:2008jb, Kumar:2009pr, Kumar:2010mv,
Lee:2013zca, Lee:2013ega, Koga:2019mee, Agrawal:2022hnf, Blasi:2022woz, Blasi:2023rqi} for topological defects. 
}


\medskip\noindent\textit{Acknowledgments.}\,---\,
The authors are thankful to Simone Blasi for helpful comments.
JK is supported by the JSPS Overseas Research Fellowships.
JK and MY acknowledge the hospitality at Jeonbuk National University where this work was initiated during ``Korea - Japan joint workshop on Particle Physics, Cosmology, and Gravity''.
This work was supported by JSPS KAKENHI Grant Numbers 23K19048 (RJ), 20H05851 (MY) and 23K13092 (MY).
MY was supported by MEXT Leading Initiative for Excellent Young Researchers. 


\bibliographystyle{JHEP}
\bibliography{ref}

\providecommand{\href}[2]{#2}\begingroup\raggedright\begin{thebibliography}{100}

\bibitem{Sher:1988mj}
M.~Sher, \emph{{Electroweak Higgs Potentials and Vacuum Stability}},
  \href{https://doi.org/10.1016/0370-1573(89)90061-6}{\emph{Phys. Rept.}
  {\bfseries 179} (1989) 273}.

\bibitem{Arnold:1989cb}
P.~B. Arnold, \emph{{Can the Electroweak Vacuum Be Unstable?}},
  \href{https://doi.org/10.1103/PhysRevD.40.613}{\emph{Phys. Rev. D} {\bfseries
  40} (1989) 613}.

\bibitem{Altarelli:1994rb}
G.~Altarelli and G.~Isidori, \emph{{Lower limit on the Higgs mass in the
  standard model: An Update}},
  \href{https://doi.org/10.1016/0370-2693(94)91458-3}{\emph{Phys. Lett. B}
  {\bfseries 337} (1994) 141}.

\bibitem{Espinosa:1995se}
J.~R. Espinosa and M.~Quiros, \emph{{Improved metastability bounds on the
  standard model Higgs mass}},
  \href{https://doi.org/10.1016/0370-2693(95)00572-3}{\emph{Phys. Lett. B}
  {\bfseries 353} (1995) 257}
  [\href{https://arxiv.org/abs/hep-ph/9504241}{{\ttfamily hep-ph/9504241}}].

\bibitem{Casas:1996aq}
J.~A. Casas, J.~R. Espinosa and M.~Quiros, \emph{{Standard model stability
  bounds for new physics within LHC reach}},
  \href{https://doi.org/10.1016/0370-2693(96)00682-X}{\emph{Phys. Lett. B}
  {\bfseries 382} (1996) 374}
  [\href{https://arxiv.org/abs/hep-ph/9603227}{{\ttfamily hep-ph/9603227}}].

\bibitem{Hambye:1996wb}
T.~Hambye and K.~Riesselmann, \emph{{Matching conditions and Higgs mass upper
  bounds revisited}},
  \href{https://doi.org/10.1103/PhysRevD.55.7255}{\emph{Phys. Rev. D}
  {\bfseries 55} (1997) 7255}
  [\href{https://arxiv.org/abs/hep-ph/9610272}{{\ttfamily hep-ph/9610272}}].

\bibitem{Isidori:2001bm}
G.~Isidori, G.~Ridolfi and A.~Strumia, \emph{{On the metastability of the
  standard model vacuum}},
  \href{https://doi.org/10.1016/S0550-3213(01)00302-9}{\emph{Nucl. Phys. B}
  {\bfseries 609} (2001) 387}
  [\href{https://arxiv.org/abs/hep-ph/0104016}{{\ttfamily hep-ph/0104016}}].

\bibitem{Espinosa:2007qp}
J.~R. Espinosa, G.~F. Giudice and A.~Riotto, \emph{{Cosmological implications
  of the Higgs mass measurement}},
  \href{https://doi.org/10.1088/1475-7516/2008/05/002}{\emph{JCAP} {\bfseries
  05} (2008) 002} [\href{https://arxiv.org/abs/0710.2484}{{\ttfamily
  0710.2484}}].

\bibitem{Ellis:2009tp}
J.~Ellis, J.~R. Espinosa, G.~F. Giudice, A.~Hoecker and A.~Riotto, \emph{{The
  Probable Fate of the Standard Model}},
  \href{https://doi.org/10.1016/j.physletb.2009.07.054}{\emph{Phys. Lett. B}
  {\bfseries 679} (2009) 369}
  [\href{https://arxiv.org/abs/0906.0954}{{\ttfamily 0906.0954}}].

\bibitem{Bezrukov:2012sa}
F.~Bezrukov, M.~Y. Kalmykov, B.~A. Kniehl and M.~Shaposhnikov, \emph{{Higgs
  Boson Mass and New Physics}},
  \href{https://doi.org/10.1007/JHEP10(2012)140}{\emph{JHEP} {\bfseries 10}
  (2012) 140} [\href{https://arxiv.org/abs/1205.2893}{{\ttfamily 1205.2893}}].

\bibitem{Bednyakov:2015sca}
A.~V. Bednyakov, B.~A. Kniehl, A.~F. Pikelner and O.~L. Veretin,
  \emph{{Stability of the Electroweak Vacuum: Gauge Independence and Advanced
  Precision}},
  \href{https://doi.org/10.1103/PhysRevLett.115.201802}{\emph{Phys. Rev. Lett.}
  {\bfseries 115} (2015) 201802}
  [\href{https://arxiv.org/abs/1507.08833}{{\ttfamily 1507.08833}}].

\bibitem{EliasMiro:2011aa}
J.~Elias-Miro, J.~R. Espinosa, G.~F. Giudice, G.~Isidori, A.~Riotto and
  A.~Strumia, \emph{{Higgs mass implications on the stability of the
  electroweak vacuum}},
  \href{https://doi.org/10.1016/j.physletb.2012.02.013}{\emph{Phys. Lett. B}
  {\bfseries 709} (2012) 222}
  [\href{https://arxiv.org/abs/1112.3022}{{\ttfamily 1112.3022}}].

\bibitem{Degrassi:2012ry}
G.~Degrassi, S.~Di~Vita, J.~Elias-Miro, J.~R. Espinosa, G.~F. Giudice,
  G.~Isidori et~al., \emph{{Higgs mass and vacuum stability in the Standard
  Model at NNLO}}, \href{https://doi.org/10.1007/JHEP08(2012)098}{\emph{JHEP}
  {\bfseries 08} (2012) 098} [\href{https://arxiv.org/abs/1205.6497}{{\ttfamily
  1205.6497}}].

\bibitem{Buttazzo:2013uya}
D.~Buttazzo, G.~Degrassi, P.~P. Giardino, G.~F. Giudice, F.~Sala, A.~Salvio
  et~al., \emph{{Investigating the near-criticality of the Higgs boson}},
  \href{https://doi.org/10.1007/JHEP12(2013)089}{\emph{JHEP} {\bfseries 12}
  (2013) 089} [\href{https://arxiv.org/abs/1307.3536}{{\ttfamily 1307.3536}}].

\bibitem{Branchina:2013jra}
V.~Branchina and E.~Messina, \emph{{Stability, Higgs Boson Mass and New
  Physics}}, \href{https://doi.org/10.1103/PhysRevLett.111.241801}{\emph{Phys.
  Rev. Lett.} {\bfseries 111} (2013) 241801}
  [\href{https://arxiv.org/abs/1307.5193}{{\ttfamily 1307.5193}}].

\bibitem{Markkanen:2018pdo}
T.~Markkanen, A.~Rajantie and S.~Stopyra, \emph{{Cosmological Aspects of Higgs
  Vacuum Metastability}},
  \href{https://doi.org/10.3389/fspas.2018.00040}{\emph{Front. Astron. Space
  Sci.} {\bfseries 5} (2018) 40}
  [\href{https://arxiv.org/abs/1809.06923}{{\ttfamily 1809.06923}}].

\bibitem{Hiscock:1987hn}
W.~A. Hiscock, \emph{{CAN BLACK HOLES NUCLEATE VACUUM PHASE TRANSITIONS?}},
  \href{https://doi.org/10.1103/PhysRevD.35.1161}{\emph{Phys. Rev. D}
  {\bfseries 35} (1987) 1161}.

\bibitem{Berezin:1987ea}
V.~A. Berezin, V.~A. Kuzmin and I.~I. Tkachev, \emph{{O(3) Invariant Tunneling
  in General Relativity}},
  \href{https://doi.org/10.1016/0370-2693(88)90672-7}{\emph{Phys. Lett. B}
  {\bfseries 207} (1988) 397}.

\bibitem{Arnold:1989cq}
P.~B. Arnold, \emph{{GRAVITY AND FALSE VACUUM DECAY RATES: O(3) SOLUTIONS}},
  \href{https://doi.org/10.1016/0550-3213(90)90243-7}{\emph{Nucl. Phys. B}
  {\bfseries 346} (1990) 160}.

\bibitem{Berezin:1990qs}
V.~A. Berezin, V.~A. Kuzmin and I.~I. Tkachev, \emph{{Black holes initiate
  false vacuum decay}},
  \href{https://doi.org/10.1103/PhysRevD.43.R3112}{\emph{Phys. Rev. D}
  {\bfseries 43} (1991) 3112}.

\bibitem{Samuel:1992wt}
D.~A. Samuel and W.~A. Hiscock, \emph{{Gravitationally compact objects as
  nucleation sites for first order vacuum phase transitions}},
  \href{https://doi.org/10.1103/PhysRevD.45.4411}{\emph{Phys. Rev. D}
  {\bfseries 45} (1992) 4411}.

\bibitem{Gomberoff:2003zh}
A.~Gomberoff, M.~Henneaux, C.~Teitelboim and F.~Wilczek, \emph{{Thermal decay
  of the cosmological constant into black holes}},
  \href{https://doi.org/10.1103/PhysRevD.69.083520}{\emph{Phys. Rev. D}
  {\bfseries 69} (2004) 083520}
  [\href{https://arxiv.org/abs/hep-th/0311011}{{\ttfamily hep-th/0311011}}].

\bibitem{Garriga:2004nm}
J.~Garriga and A.~Megevand, \emph{{Decay of de Sitter vacua by thermal
  activation}},
  \href{https://doi.org/10.1023/B:IJTP.0000048178.69097.fb}{\emph{Int. J.
  Theor. Phys.} {\bfseries 43} (2004) 883}
  [\href{https://arxiv.org/abs/hep-th/0404097}{{\ttfamily hep-th/0404097}}].

\bibitem{Gregory:2013hja}
R.~Gregory, I.~G. Moss and B.~Withers, \emph{{Black holes as bubble nucleation
  sites}}, \href{https://doi.org/10.1007/JHEP03(2014)081}{\emph{JHEP}
  {\bfseries 03} (2014) 081} [\href{https://arxiv.org/abs/1401.0017}{{\ttfamily
  1401.0017}}].

\bibitem{Burda:2015isa}
P.~Burda, R.~Gregory and I.~Moss, \emph{{Gravity and the stability of the Higgs
  vacuum}}, \href{https://doi.org/10.1103/PhysRevLett.115.071303}{\emph{Phys.
  Rev. Lett.} {\bfseries 115} (2015) 071303}
  [\href{https://arxiv.org/abs/1501.04937}{{\ttfamily 1501.04937}}].

\bibitem{Burda:2015yfa}
P.~Burda, R.~Gregory and I.~Moss, \emph{{Vacuum metastability with black
  holes}}, \href{https://doi.org/10.1007/JHEP08(2015)114}{\emph{JHEP}
  {\bfseries 08} (2015) 114}
  [\href{https://arxiv.org/abs/1503.07331}{{\ttfamily 1503.07331}}].

\bibitem{Burda:2016mou}
P.~Burda, R.~Gregory and I.~Moss, \emph{{The fate of the Higgs vacuum}},
  \href{https://doi.org/10.1007/JHEP06(2016)025}{\emph{JHEP} {\bfseries 06}
  (2016) 025} [\href{https://arxiv.org/abs/1601.02152}{{\ttfamily
  1601.02152}}].

\bibitem{Chen:2017suz}
P.~Chen, G.~Dom\`enech, M.~Sasaki and D.-h. Yeom, \emph{{Thermal activation of
  thin-shells in anti-de Sitter black hole spacetime}},
  \href{https://doi.org/10.1007/JHEP07(2017)134}{\emph{JHEP} {\bfseries 07}
  (2017) 134} [\href{https://arxiv.org/abs/1704.04020}{{\ttfamily
  1704.04020}}].

\bibitem{Gorbunov:2017fhq}
D.~Gorbunov, D.~Levkov and A.~Panin, \emph{{Fatal youth of the Universe: black
  hole threat for the electroweak vacuum during preheating}},
  \href{https://doi.org/10.1088/1475-7516/2017/10/016}{\emph{JCAP} {\bfseries
  10} (2017) 016} [\href{https://arxiv.org/abs/1704.05399}{{\ttfamily
  1704.05399}}].

\bibitem{Canko:2017ebb}
D.~Canko, I.~Gialamas, G.~Jelic-Cizmek, A.~Riotto and N.~Tetradis, \emph{{On
  the Catalysis of the Electroweak Vacuum Decay by Black Holes at High
  Temperature}},
  \href{https://doi.org/10.1140/epjc/s10052-018-5808-y}{\emph{Eur. Phys. J. C}
  {\bfseries 78} (2018) 328}
  [\href{https://arxiv.org/abs/1706.01364}{{\ttfamily 1706.01364}}].

\bibitem{Mukaida:2017bgd}
K.~Mukaida and M.~Yamada, \emph{{False Vacuum Decay Catalyzed by Black Holes}},
  \href{https://doi.org/10.1103/PhysRevD.96.103514}{\emph{Phys. Rev. D}
  {\bfseries 96} (2017) 103514}
  [\href{https://arxiv.org/abs/1706.04523}{{\ttfamily 1706.04523}}].

\bibitem{Kohri:2017ybt}
K.~Kohri and H.~Matsui, \emph{{Electroweak Vacuum Collapse induced by Vacuum
  Fluctuations of the Higgs Field around Evaporating Black Holes}},
  \href{https://doi.org/10.1103/PhysRevD.98.123509}{\emph{Phys. Rev. D}
  {\bfseries 98} (2018) 123509}
  [\href{https://arxiv.org/abs/1708.02138}{{\ttfamily 1708.02138}}].

\bibitem{Billam:2018pvp}
T.~P. Billam, R.~Gregory, F.~Michel and I.~G. Moss, \emph{{Simulating seeded
  vacuum decay in a cold atom system}},
  \href{https://doi.org/10.1103/PhysRevD.100.065016}{\emph{Phys. Rev. D}
  {\bfseries 100} (2019) 065016}
  [\href{https://arxiv.org/abs/1811.09169}{{\ttfamily 1811.09169}}].

\bibitem{Oshita:2019jan}
N.~Oshita, K.~Ueda and M.~Yamaguchi, \emph{{Vacuum decays around spinning black
  holes}}, \href{https://doi.org/10.1007/JHEP01(2020)015}{\emph{JHEP}
  {\bfseries 01} (2020) 015}
  [\href{https://arxiv.org/abs/1909.01378}{{\ttfamily 1909.01378}}].

\bibitem{Gregory:2020cvy}
R.~Gregory, I.~G. Moss and N.~Oshita, \emph{{Black Holes, Oscillating
  Instantons, and the Hawking-Moss transition}},
  \href{https://doi.org/10.1007/JHEP07(2020)024}{\emph{JHEP} {\bfseries 07}
  (2020) 024} [\href{https://arxiv.org/abs/2003.04927}{{\ttfamily
  2003.04927}}].

\bibitem{Hayashi:2020ocn}
T.~Hayashi, K.~Kamada, N.~Oshita and J.~Yokoyama, \emph{{On catalyzed vacuum
  decay around a radiating black hole and the crisis of the electroweak
  vacuum}}, \href{https://doi.org/10.1007/JHEP08(2020)088}{\emph{JHEP}
  {\bfseries 08} (2020) 088}
  [\href{https://arxiv.org/abs/2005.12808}{{\ttfamily 2005.12808}}].

\bibitem{El-Menoufi:2020ron}
B.~K. El-Menoufi, S.~J. Huber and J.~P. Manuel, \emph{{Black holes seeding
  cosmological phase transitions}},
  \href{https://arxiv.org/abs/2006.16275}{{\ttfamily 2006.16275}}.

\bibitem{Gregory:2020hia}
R.~Gregory, I.~G. Moss, N.~Oshita and S.~Patrick, \emph{{Hawking-Moss
  transition with a black hole seed}},
  \href{https://doi.org/10.1007/JHEP09(2020)135}{\emph{JHEP} {\bfseries 09}
  (2020) 135} [\href{https://arxiv.org/abs/2007.11428}{{\ttfamily
  2007.11428}}].

\bibitem{Shkerin:2021zbf}
A.~Shkerin and S.~Sibiryakov, \emph{{Black hole induced false vacuum decay from
  first principles}},
  \href{https://doi.org/10.1007/JHEP11(2021)197}{\emph{JHEP} {\bfseries 11}
  (2021) 197} [\href{https://arxiv.org/abs/2105.09331}{{\ttfamily
  2105.09331}}].

\bibitem{Shkerin:2021rhy}
A.~Shkerin and S.~Sibiryakov, \emph{{Black hole induced false vacuum decay: the
  role of greybody factors}},
  \href{https://doi.org/10.1007/JHEP08(2022)161}{\emph{JHEP} {\bfseries 08}
  (2022) 161} [\href{https://arxiv.org/abs/2111.08017}{{\ttfamily
  2111.08017}}].

\bibitem{Strumia:2022jil}
A.~Strumia, \emph{{Black holes don\textquoteright{}t source fast Higgs vacuum
  decay}}, \href{https://doi.org/10.1007/JHEP03(2023)039}{\emph{JHEP}
  {\bfseries 03} (2023) 039}
  [\href{https://arxiv.org/abs/2209.05504}{{\ttfamily 2209.05504}}].

\bibitem{Briaud:2022few}
V.~Briaud, A.~Shkerin and S.~Sibiryakov, \emph{{Thermal false vacuum decay
  around black holes}},
  \href{https://doi.org/10.1103/PhysRevD.106.125001}{\emph{Phys. Rev. D}
  {\bfseries 106} (2022) 125001}
  [\href{https://arxiv.org/abs/2210.08028}{{\ttfamily 2210.08028}}].

\bibitem{Gregory:2023kam}
R.~Gregory and S.-Q. Hu, \emph{{Seeded vacuum decay with Gauss-Bonnet}},
  \href{https://arxiv.org/abs/2305.03006}{{\ttfamily 2305.03006}}.

\bibitem{Oshita:2018ptr}
N.~Oshita, M.~Yamada and M.~Yamaguchi, \emph{{Compact objects as the catalysts
  for vacuum decays}},
  \href{https://doi.org/10.1016/j.physletb.2019.02.032}{\emph{Phys. Lett. B}
  {\bfseries 791} (2019) 149}
  [\href{https://arxiv.org/abs/1808.01382}{{\ttfamily 1808.01382}}].

\bibitem{Kosowsky:1991ua}
A.~Kosowsky, M.~S. Turner and R.~Watkins, \emph{{Gravitational radiation from
  colliding vacuum bubbles}},
  \href{https://doi.org/10.1103/PhysRevD.45.4514}{\emph{Phys. Rev. D}
  {\bfseries 45} (1992) 4514}.

\bibitem{Kosowsky:1992rz}
A.~Kosowsky, M.~S. Turner and R.~Watkins, \emph{{Gravitational waves from first
  order cosmological phase transitions}},
  \href{https://doi.org/10.1103/PhysRevLett.69.2026}{\emph{Phys. Rev. Lett.}
  {\bfseries 69} (1992) 2026}.

\bibitem{Kosowsky:1992vn}
A.~Kosowsky and M.~S. Turner, \emph{{Gravitational radiation from colliding
  vacuum bubbles: envelope approximation to many bubble collisions}},
  \href{https://doi.org/10.1103/PhysRevD.47.4372}{\emph{Phys. Rev. D}
  {\bfseries 47} (1993) 4372}
  [\href{https://arxiv.org/abs/astro-ph/9211004}{{\ttfamily
  astro-ph/9211004}}].

\bibitem{Kamionkowski:1993fg}
M.~Kamionkowski, A.~Kosowsky and M.~S. Turner, \emph{{Gravitational radiation
  from first order phase transitions}},
  \href{https://doi.org/10.1103/PhysRevD.49.2837}{\emph{Phys. Rev. D}
  {\bfseries 49} (1994) 2837}
  [\href{https://arxiv.org/abs/astro-ph/9310044}{{\ttfamily
  astro-ph/9310044}}].

\bibitem{Huber:2008hg}
S.~J. Huber and T.~Konstandin, \emph{{Gravitational Wave Production by
  Collisions: More Bubbles}},
  \href{https://doi.org/10.1088/1475-7516/2008/09/022}{\emph{JCAP} {\bfseries
  09} (2008) 022} [\href{https://arxiv.org/abs/0806.1828}{{\ttfamily
  0806.1828}}].

\bibitem{Jinno:2016vai}
R.~Jinno and M.~Takimoto, \emph{{Gravitational waves from bubble collisions: An
  analytic derivation}},
  \href{https://doi.org/10.1103/PhysRevD.95.024009}{\emph{Phys. Rev. D}
  {\bfseries 95} (2017) 024009}
  [\href{https://arxiv.org/abs/1605.01403}{{\ttfamily 1605.01403}}].

\bibitem{Jinno:2017fby}
R.~Jinno and M.~Takimoto, \emph{{Gravitational waves from bubble dynamics:
  Beyond the Envelope}},
  \href{https://doi.org/10.1088/1475-7516/2019/01/060}{\emph{JCAP} {\bfseries
  01} (2019) 060} [\href{https://arxiv.org/abs/1707.03111}{{\ttfamily
  1707.03111}}].

\bibitem{Kodama:1981gu}
H.~Kodama, M.~Sasaki, K.~Sato and K.-i. Maeda, \emph{{Fate of Wormholes Created
  by First Order Phase Transition in the Early Universe}},
  \href{https://doi.org/10.1143/PTP.66.2052}{\emph{Prog. Theor. Phys.}
  {\bfseries 66} (1981) 2052}.

\bibitem{Berezin:1982ur}
V.~A. Berezin, V.~A. Kuzmin and I.~I. Tkachev, \emph{{THIN WALL VACUUM DOMAINS
  EVOLUTION}}, \href{https://doi.org/10.1016/0370-2693(83)90630-5}{\emph{Phys.
  Lett. B} {\bfseries 120} (1983) 91}.

\bibitem{Blau:1986cw}
S.~K. Blau, E.~I. Guendelman and A.~H. Guth, \emph{{The Dynamics of False
  Vacuum Bubbles}}, \href{https://doi.org/10.1103/PhysRevD.35.1747}{\emph{Phys.
  Rev. D} {\bfseries 35} (1987) 1747}.

\bibitem{Maeda:2009tk}
H.~Maeda, T.~Harada and B.~J. Carr, \emph{{Cosmological wormholes}},
  \href{https://doi.org/10.1103/PhysRevD.79.044034}{\emph{Phys. Rev. D}
  {\bfseries 79} (2009) 044034}
  [\href{https://arxiv.org/abs/0901.1153}{{\ttfamily 0901.1153}}].

\bibitem{Garriga:2015fdk}
J.~Garriga, A.~Vilenkin and J.~Zhang, \emph{{Black holes and the multiverse}},
  \href{https://doi.org/10.1088/1475-7516/2016/02/064}{\emph{JCAP} {\bfseries
  02} (2016) 064} [\href{https://arxiv.org/abs/1512.01819}{{\ttfamily
  1512.01819}}].

\bibitem{Deng:2016vzb}
H.~Deng, J.~Garriga and A.~Vilenkin, \emph{{Primordial black hole and wormhole
  formation by domain walls}},
  \href{https://doi.org/10.1088/1475-7516/2017/04/050}{\emph{JCAP} {\bfseries
  04} (2017) 050} [\href{https://arxiv.org/abs/1612.03753}{{\ttfamily
  1612.03753}}].

\bibitem{Deng:2017uwc}
H.~Deng and A.~Vilenkin, \emph{{Primordial black hole formation by vacuum
  bubbles}}, \href{https://doi.org/10.1088/1475-7516/2017/12/044}{\emph{JCAP}
  {\bfseries 12} (2017) 044}
  [\href{https://arxiv.org/abs/1710.02865}{{\ttfamily 1710.02865}}].

\bibitem{Deng:2018cxb}
H.~Deng, A.~Vilenkin and M.~Yamada, \emph{{CMB spectral distortions from black
  holes formed by vacuum bubbles}},
  \href{https://doi.org/10.1088/1475-7516/2018/07/059}{\emph{JCAP} {\bfseries
  07} (2018) 059} [\href{https://arxiv.org/abs/1804.10059}{{\ttfamily
  1804.10059}}].

\bibitem{Sato:1981bf}
K.~Sato, M.~Sasaki, H.~Kodama and K.-i. Maeda, \emph{{Creation of Wormholes by
  First Order Phase Transition of a Vacuum in the Early Universe}},
  \href{https://doi.org/10.1143/PTP.65.1443}{\emph{Prog. Theor. Phys.}
  {\bfseries 65} (1981) 1443}.

\bibitem{Maeda:1981gw}
K.-i. Maeda, K.~Sato, M.~Sasaki and H.~Kodama, \emph{{Creation of De
  Sitter-schwarzschild Wormholes by a Cosmological First Order Phase
  Transition}}, \href{https://doi.org/10.1016/0370-2693(82)91151-0}{\emph{Phys.
  Lett. B} {\bfseries 108} (1982) 98}.

\bibitem{Sato:1981gv}
K.~Sato, H.~Kodama, M.~Sasaki and K.-i. Maeda, \emph{{Multiproduction of
  Universes by First Order Phase Transition of a Vacuum}},
  \href{https://doi.org/10.1016/0370-2693(82)91152-2}{\emph{Phys. Lett. B}
  {\bfseries 108} (1982) 103}.

\bibitem{Kodama:1982sf}
H.~Kodama, M.~Sasaki and K.~Sato, \emph{{Abundance of Primordial Holes Produced
  by Cosmological First Order Phase Transition}},
  \href{https://doi.org/10.1143/PTP.68.1979}{\emph{Prog. Theor. Phys.}
  {\bfseries 68} (1982) 1979}.

\bibitem{Hawking:1982ga}
S.~W. Hawking, I.~G. Moss and J.~M. Stewart, \emph{{Bubble Collisions in the
  Very Early Universe}},
  \href{https://doi.org/10.1103/PhysRevD.26.2681}{\emph{Phys. Rev. D}
  {\bfseries 26} (1982) 2681}.

\bibitem{Moss:1994iq}
I.~G. Moss, \emph{{Singularity formation from colliding bubbles}},
  \href{https://doi.org/10.1103/PhysRevD.50.676}{\emph{Phys. Rev. D} {\bfseries
  50} (1994) 676}.

\bibitem{Khlopov:1998nm}
M.~Y. Khlopov, R.~V. Konoplich, S.~G. Rubin and A.~S. Sakharov,
  \emph{{Formation of black holes in first order phase transitions}},
  \href{https://arxiv.org/abs/hep-ph/9807343}{{\ttfamily hep-ph/9807343}}.

\bibitem{Jedamzik:1999am}
K.~Jedamzik and J.~C. Niemeyer, \emph{{Primordial black hole formation during
  first order phase transitions}},
  \href{https://doi.org/10.1103/PhysRevD.59.124014}{\emph{Phys. Rev. D}
  {\bfseries 59} (1999) 124014}
  [\href{https://arxiv.org/abs/astro-ph/9901293}{{\ttfamily
  astro-ph/9901293}}].

\bibitem{Lewicki:2019gmv}
M.~Lewicki and V.~Vaskonen, \emph{{On bubble collisions in strongly supercooled
  phase transitions}},
  \href{https://doi.org/10.1016/j.dark.2020.100672}{\emph{Phys. Dark Univ.}
  {\bfseries 30} (2020) 100672}
  [\href{https://arxiv.org/abs/1912.00997}{{\ttfamily 1912.00997}}].

\bibitem{Gross:2021qgx}
C.~Gross, G.~Landini, A.~Strumia and D.~Teresi, \emph{{Dark Matter as dark
  dwarfs and other macroscopic objects: multiverse relics?}},
  \href{https://doi.org/10.1007/JHEP09(2021)033}{\emph{JHEP} {\bfseries 09}
  (2021) 033} [\href{https://arxiv.org/abs/2105.02840}{{\ttfamily
  2105.02840}}].

\bibitem{Baker:2021nyl}
M.~J. Baker, M.~Breitbach, J.~Kopp and L.~Mittnacht, \emph{{Primordial Black
  Holes from First-Order Cosmological Phase Transitions}},
  \href{https://arxiv.org/abs/2105.07481}{{\ttfamily 2105.07481}}.

\bibitem{Kawana:2021tde}
K.~Kawana and K.-P. Xie, \emph{{Primordial black holes from a cosmic phase
  transition: The collapse of Fermi-balls}},
  \href{https://doi.org/10.1016/j.physletb.2021.136791}{\emph{Phys. Lett. B}
  {\bfseries 824} (2022) 136791}
  [\href{https://arxiv.org/abs/2106.00111}{{\ttfamily 2106.00111}}].

\bibitem{Liu:2021svg}
J.~Liu, L.~Bian, R.-G. Cai, Z.-K. Guo and S.-J. Wang, \emph{{Primordial black
  hole production during first-order phase transitions}},
  \href{https://doi.org/10.1103/PhysRevD.105.L021303}{\emph{Phys. Rev. D}
  {\bfseries 105} (2022) L021303}
  [\href{https://arxiv.org/abs/2106.05637}{{\ttfamily 2106.05637}}].

\bibitem{Baker:2021sno}
M.~J. Baker, M.~Breitbach, J.~Kopp and L.~Mittnacht, \emph{{Detailed
  Calculation of Primordial Black Hole Formation During First-Order
  Cosmological Phase Transitions}},
  \href{https://arxiv.org/abs/2110.00005}{{\ttfamily 2110.00005}}.

\bibitem{Jung:2021mku}
T.~H. Jung and T.~Okui, \emph{{Primordial black holes from bubble collisions
  during a first-order phase transition}},
  \href{https://arxiv.org/abs/2110.04271}{{\ttfamily 2110.04271}}.

\bibitem{Hashino:2021qoq}
K.~Hashino, S.~Kanemura and T.~Takahashi, \emph{{Primordial black holes as a
  probe of strongly first-order electroweak phase transition}},
  \href{https://doi.org/10.1016/j.physletb.2022.137261}{\emph{Phys. Lett. B}
  {\bfseries 833} (2022) 137261}
  [\href{https://arxiv.org/abs/2111.13099}{{\ttfamily 2111.13099}}].

\bibitem{Huang:2022him}
P.~Huang and K.-P. Xie, \emph{{Primordial black holes from an electroweak phase
  transition}}, \href{https://doi.org/10.1103/PhysRevD.105.115033}{\emph{Phys.
  Rev. D} {\bfseries 105} (2022) 115033}
  [\href{https://arxiv.org/abs/2201.07243}{{\ttfamily 2201.07243}}].

\bibitem{He:2022amv}
S.~He, L.~Li, Z.~Li and S.-J. Wang, \emph{{Gravitational Waves and Primordial
  Black Hole Productions from Gluodynamics}},
  \href{https://arxiv.org/abs/2210.14094}{{\ttfamily 2210.14094}}.

\bibitem{Hashino:2022tcs}
K.~Hashino, S.~Kanemura, T.~Takahashi and M.~Tanaka, \emph{{Probing first-order
  electroweak phase transition via primordial black holes in the effective
  field theory}},
  \href{https://doi.org/10.1016/j.physletb.2023.137688}{\emph{Phys. Lett. B}
  {\bfseries 838} (2023) 137688}
  [\href{https://arxiv.org/abs/2211.16225}{{\ttfamily 2211.16225}}].

\bibitem{Kawana:2022olo}
K.~Kawana, T.~Kim and P.~Lu, \emph{{PBH Formation from Overdensities in Delayed
  Vacuum Transitions}},  \href{https://arxiv.org/abs/2212.14037}{{\ttfamily
  2212.14037}}.

\bibitem{Lewicki:2023ioy}
M.~Lewicki, P.~Toczek and V.~Vaskonen, \emph{{Primordial black holes from
  strong first-order phase transitions}},
  \href{https://arxiv.org/abs/2305.04924}{{\ttfamily 2305.04924}}.

\bibitem{Gouttenoire:2023naa}
Y.~Gouttenoire and T.~Volansky, \emph{{Primordial Black Holes from Supercooled
  Phase Transitions}},  \href{https://arxiv.org/abs/2305.04942}{{\ttfamily
  2305.04942}}.

\bibitem{Salvio:2023ynn}
A.~Salvio, \emph{{Supercooling in Radiative Symmetry Breaking: Theory
  Extensions, Gravitational Wave Detection and Primordial Black Holes}},
  \href{https://arxiv.org/abs/2307.04694}{{\ttfamily 2307.04694}}.

\bibitem{Baldes:2023rqv}
I.~Baldes and M.~O. Olea-Romacho, \emph{{Primordial black holes as dark matter:
  Interferometric tests of phase transition origin}},
  \href{https://arxiv.org/abs/2307.11639}{{\ttfamily 2307.11639}}.

\bibitem{Carr:1975qj}
B.~J. Carr, \emph{{The Primordial black hole mass spectrum}},
  \href{https://doi.org/10.1086/153853}{\emph{Astrophys. J.} {\bfseries 201}
  (1975) 1}.

\bibitem{Megevand:2016lpr}
A.~Megevand and S.~Ramirez, \emph{{Bubble nucleation and growth in very strong
  cosmological phase transitions}},
  \href{https://doi.org/10.1016/j.nuclphysb.2017.03.009}{\emph{Nucl. Phys. B}
  {\bfseries 919} (2017) 74}
  [\href{https://arxiv.org/abs/1611.05853}{{\ttfamily 1611.05853}}].

\bibitem{Kobakhidze:2017mru}
A.~Kobakhidze, C.~Lagger, A.~Manning and J.~Yue, \emph{{Gravitational waves
  from a supercooled electroweak phase transition and their detection with
  pulsar timing arrays}},
  \href{https://doi.org/10.1140/epjc/s10052-017-5132-y}{\emph{Eur. Phys. J. C}
  {\bfseries 77} (2017) 570}
  [\href{https://arxiv.org/abs/1703.06552}{{\ttfamily 1703.06552}}].

\bibitem{Cai:2017tmh}
R.-G. Cai, M.~Sasaki and S.-J. Wang, \emph{{The gravitational waves from the
  first-order phase transition with a dimension-six operator}},
  \href{https://doi.org/10.1088/1475-7516/2017/08/004}{\emph{JCAP} {\bfseries
  08} (2017) 004} [\href{https://arxiv.org/abs/1707.03001}{{\ttfamily
  1707.03001}}].

\bibitem{Megevand:2017vtb}
A.~M\'egevand and S.~Ram\'\i{}rez, \emph{{Bubble nucleation and growth in slow
  cosmological phase transitions}},
  \href{https://doi.org/10.1016/j.nuclphysb.2018.01.012}{\emph{Nucl. Phys. B}
  {\bfseries 928} (2018) 38}
  [\href{https://arxiv.org/abs/1710.06279}{{\ttfamily 1710.06279}}].

\bibitem{Ellis:2018mja}
J.~Ellis, M.~Lewicki and J.~M. No, \emph{{On the Maximal Strength of a
  First-Order Electroweak Phase Transition and its Gravitational Wave Signal}},
  \href{https://doi.org/10.1088/1475-7516/2019/04/003}{\emph{JCAP} {\bfseries
  04} (2019) 003} [\href{https://arxiv.org/abs/1809.08242}{{\ttfamily
  1809.08242}}].

\bibitem{Wang:2020jrd}
X.~Wang, F.~P. Huang and X.~Zhang, \emph{{Phase transition dynamics and
  gravitational wave spectra of strong first-order phase transition in
  supercooled universe}},
  \href{https://doi.org/10.1088/1475-7516/2020/05/045}{\emph{JCAP} {\bfseries
  05} (2020) 045} [\href{https://arxiv.org/abs/2003.08892}{{\ttfamily
  2003.08892}}].

\bibitem{Athron:2022mmm}
P.~Athron, C.~Bal\'azs and L.~Morris, \emph{{Supercool subtleties of
  cosmological phase transitions}},
  \href{https://doi.org/10.1088/1475-7516/2023/03/006}{\emph{JCAP} {\bfseries
  03} (2023) 006} [\href{https://arxiv.org/abs/2212.07559}{{\ttfamily
  2212.07559}}].

\bibitem{Athron:2023mer}
P.~Athron, A.~Fowlie, C.-T. Lu, L.~Morris, L.~Wu, Y.~Wu et~al., \emph{{Can
  supercooled phase transitions explain the gravitational wave background
  observed by pulsar timing arrays?}},
  \href{https://arxiv.org/abs/2306.17239}{{\ttfamily 2306.17239}}.

\bibitem{Deng:2020mds}
H.~Deng, \emph{{Primordial black hole formation by vacuum bubbles. Part II}},
  \href{https://doi.org/10.1088/1475-7516/2020/09/023}{\emph{JCAP} {\bfseries
  09} (2020) 023} [\href{https://arxiv.org/abs/2006.11907}{{\ttfamily
  2006.11907}}].

\bibitem{Carr:2020gox}
B.~Carr, K.~Kohri, Y.~Sendouda and J.~Yokoyama, \emph{{Constraints on
  primordial black holes}},
  \href{https://doi.org/10.1088/1361-6633/ac1e31}{\emph{Rept. Prog. Phys.}
  {\bfseries 84} (2021) 116902}
  [\href{https://arxiv.org/abs/2002.12778}{{\ttfamily 2002.12778}}].

\bibitem{Caprini:2015zlo}
C.~Caprini et~al., \emph{{Science with the space-based interferometer eLISA.
  II: Gravitational waves from cosmological phase transitions}},
  \href{https://doi.org/10.1088/1475-7516/2016/04/001}{\emph{JCAP} {\bfseries
  04} (2016) 001} [\href{https://arxiv.org/abs/1512.06239}{{\ttfamily
  1512.06239}}].

\bibitem{Manchester:2012za}
R.~N. Manchester et~al., \emph{{The Parkes Pulsar Timing Array Project}},
  \href{https://doi.org/10.1017/pasa.2012.017}{\emph{Publ. Astron. Soc.
  Austral.} {\bfseries 30} (2013) 17}
  [\href{https://arxiv.org/abs/1210.6130}{{\ttfamily 1210.6130}}].

\bibitem{McLaughlin:2013ira}
M.~A. McLaughlin, \emph{{The North American Nanohertz Observatory for
  Gravitational Waves}},
  \href{https://doi.org/10.1088/0264-9381/30/22/224008}{\emph{Class. Quant.
  Grav.} {\bfseries 30} (2013) 224008}
  [\href{https://arxiv.org/abs/1310.0758}{{\ttfamily 1310.0758}}].

\bibitem{Kramer:2013kea}
{\scshape EPTA} collaboration, M.~Kramer and D.~J. Champion, \emph{{The
  European Pulsar Timing Array and the Large European Array for Pulsars}},
  \href{https://doi.org/10.1088/0264-9381/30/22/224009}{\emph{Class. Quant.
  Grav.} {\bfseries 30} (2013) 224009}.

\bibitem{Manchester:2013ndt}
R.~N. Manchester, \emph{{The International Pulsar Timing Array}},
  \href{https://doi.org/10.1088/0264-9381/30/22/224010}{\emph{Class. Quant.
  Grav.} {\bfseries 30} (2013) 224010}
  [\href{https://arxiv.org/abs/1309.7392}{{\ttfamily 1309.7392}}].

\bibitem{Janssen:2014dka}
G.~Janssen et~al., \emph{{Gravitational wave astronomy with the SKA}},
  \href{https://doi.org/10.22323/1.215.0037}{\emph{PoS} {\bfseries AASKA14}
  (2015) 037} [\href{https://arxiv.org/abs/1501.00127}{{\ttfamily
  1501.00127}}].

\bibitem{Punturo:2010zz}
M.~Punturo et~al., \emph{{The Einstein Telescope: A third-generation
  gravitational wave observatory}},
  \href{https://doi.org/10.1088/0264-9381/27/19/194002}{\emph{Class. Quant.
  Grav.} {\bfseries 27} (2010) 194002}.

\bibitem{Hild:2010id}
S.~Hild et~al., \emph{{Sensitivity Studies for Third-Generation Gravitational
  Wave Observatories}},
  \href{https://doi.org/10.1088/0264-9381/28/9/094013}{\emph{Class. Quant.
  Grav.} {\bfseries 28} (2011) 094013}
  [\href{https://arxiv.org/abs/1012.0908}{{\ttfamily 1012.0908}}].

\bibitem{LIGOScientific:2016wof}
{\scshape LIGO Scientific} collaboration, B.~P. Abbott et~al., \emph{{Exploring
  the Sensitivity of Next Generation Gravitational Wave Detectors}},
  \href{https://doi.org/10.1088/1361-6382/aa51f4}{\emph{Class. Quant. Grav.}
  {\bfseries 34} (2017) 044001}
  [\href{https://arxiv.org/abs/1607.08697}{{\ttfamily 1607.08697}}].

\bibitem{Reitze:2019iox}
D.~Reitze et~al., \emph{{Cosmic Explorer: The U.S. Contribution to
  Gravitational-Wave Astronomy beyond LIGO}}, {\emph{Bull. Am. Astron. Soc.}
  {\bfseries 51} (2019) 035}
  [\href{https://arxiv.org/abs/1907.04833}{{\ttfamily 1907.04833}}].

\bibitem{https://doi.org/10.48550/arxiv.1702.00786}
P.~Amaro-Seoane, H.~Audley, S.~Babak, J.~Baker, E.~Barausse, P.~Bender et~al.,
  \emph{Laser interferometer space antenna},  2017.
\newblock 10.48550/ARXIV.1702.00786.

\bibitem{Baker:2019nia}
J.~Baker et~al., \emph{{The Laser Interferometer Space Antenna: Unveiling the
  Millihertz Gravitational Wave Sky}},
  \href{https://arxiv.org/abs/1907.06482}{{\ttfamily 1907.06482}}.

\bibitem{Seto:2001qf}
N.~Seto, S.~Kawamura and T.~Nakamura, \emph{{Possibility of direct measurement
  of the acceleration of the universe using 0.1-Hz band laser interferometer
  gravitational wave antenna in space}},
  \href{https://doi.org/10.1103/PhysRevLett.87.221103}{\emph{Phys. Rev. Lett.}
  {\bfseries 87} (2001) 221103}
  [\href{https://arxiv.org/abs/astro-ph/0108011}{{\ttfamily
  astro-ph/0108011}}].

\bibitem{Kawamura:2006up}
S.~Kawamura et~al., \emph{{The Japanese space gravitational wave antenna
  DECIGO}}, \href{https://doi.org/10.1088/0264-9381/23/8/S17}{\emph{Class.
  Quant. Grav.} {\bfseries 23} (2006) S125}.

\bibitem{Carr:2009jm}
B.~J. Carr, K.~Kohri, Y.~Sendouda and J.~Yokoyama, \emph{{New cosmological
  constraints on primordial black holes}},
  \href{https://doi.org/10.1103/PhysRevD.81.104019}{\emph{Phys. Rev. D}
  {\bfseries 81} (2010) 104019}
  [\href{https://arxiv.org/abs/0912.5297}{{\ttfamily 0912.5297}}].

\bibitem{Acharya:2020jbv}
S.~K. Acharya and R.~Khatri, \emph{{CMB and BBN constraints on evaporating
  primordial black holes revisited}},
  \href{https://doi.org/10.1088/1475-7516/2020/06/018}{\emph{JCAP} {\bfseries
  06} (2020) 018} [\href{https://arxiv.org/abs/2002.00898}{{\ttfamily
  2002.00898}}].

\bibitem{Carr:2016hva}
B.~J. Carr, K.~Kohri, Y.~Sendouda and J.~Yokoyama, \emph{{Constraints on
  primordial black holes from the Galactic gamma-ray background}},
  \href{https://doi.org/10.1103/PhysRevD.94.044029}{\emph{Phys. Rev. D}
  {\bfseries 94} (2016) 044029}
  [\href{https://arxiv.org/abs/1604.05349}{{\ttfamily 1604.05349}}].

\bibitem{Boudaud:2018hqb}
M.~Boudaud and M.~Cirelli, \emph{{Voyager 1 $e^\pm$ Further Constrain
  Primordial Black Holes as Dark Matter}},
  \href{https://doi.org/10.1103/PhysRevLett.122.041104}{\emph{Phys. Rev. Lett.}
  {\bfseries 122} (2019) 041104}
  [\href{https://arxiv.org/abs/1807.03075}{{\ttfamily 1807.03075}}].

\bibitem{Niikura:2017zjd}
H.~Niikura et~al., \emph{{Microlensing constraints on primordial black holes
  with Subaru/HSC Andromeda observations}},
  \href{https://doi.org/10.1038/s41550-019-0723-1}{\emph{Nature Astron.}
  {\bfseries 3} (2019) 524} [\href{https://arxiv.org/abs/1701.02151}{{\ttfamily
  1701.02151}}].

\bibitem{Smyth:2019whb}
N.~Smyth, S.~Profumo, S.~English, T.~Jeltema, K.~McKinnon and P.~Guhathakurta,
  \emph{{Updated Constraints on Asteroid-Mass Primordial Black Holes as Dark
  Matter}}, \href{https://doi.org/10.1103/PhysRevD.101.063005}{\emph{Phys. Rev.
  D} {\bfseries 101} (2020) 063005}
  [\href{https://arxiv.org/abs/1910.01285}{{\ttfamily 1910.01285}}].

\bibitem{Niikura:2019kqi}
H.~Niikura, M.~Takada, S.~Yokoyama, T.~Sumi and S.~Masaki, \emph{{Constraints
  on Earth-mass primordial black holes from OGLE 5-year microlensing events}},
  \href{https://doi.org/10.1103/PhysRevD.99.083503}{\emph{Phys. Rev. D}
  {\bfseries 99} (2019) 083503}
  [\href{https://arxiv.org/abs/1901.07120}{{\ttfamily 1901.07120}}].

\bibitem{EROS-2:2006ryy}
{\scshape EROS-2} collaboration, P.~Tisserand et~al., \emph{{Limits on the
  Macho Content of the Galactic Halo from the EROS-2 Survey of the Magellanic
  Clouds}}, \href{https://doi.org/10.1051/0004-6361:20066017}{\emph{Astron.
  Astrophys.} {\bfseries 469} (2007) 387}
  [\href{https://arxiv.org/abs/astro-ph/0607207}{{\ttfamily
  astro-ph/0607207}}].

\bibitem{Oguri:2017ock}
M.~Oguri, J.~M. Diego, N.~Kaiser, P.~L. Kelly and T.~Broadhurst,
  \emph{{Understanding caustic crossings in giant arcs: characteristic scales,
  event rates, and constraints on compact dark matter}},
  \href{https://doi.org/10.1103/PhysRevD.97.023518}{\emph{Phys. Rev. D}
  {\bfseries 97} (2018) 023518}
  [\href{https://arxiv.org/abs/1710.00148}{{\ttfamily 1710.00148}}].

\bibitem{Carr:2018rid}
B.~Carr and J.~Silk, \emph{{Primordial Black Holes as Generators of Cosmic
  Structures}}, \href{https://doi.org/10.1093/mnras/sty1204}{\emph{Mon. Not.
  Roy. Astron. Soc.} {\bfseries 478} (2018) 3756}
  [\href{https://arxiv.org/abs/1801.00672}{{\ttfamily 1801.00672}}].

\bibitem{Inoue:2017csr}
Y.~Inoue and A.~Kusenko, \emph{{New X-ray bound on density of primordial black
  holes}}, \href{https://doi.org/10.1088/1475-7516/2017/10/034}{\emph{JCAP}
  {\bfseries 10} (2017) 034}
  [\href{https://arxiv.org/abs/1705.00791}{{\ttfamily 1705.00791}}].

\bibitem{Serpico:2020ehh}
P.~D. Serpico, V.~Poulin, D.~Inman and K.~Kohri, \emph{{Cosmic microwave
  background bounds on primordial black holes including dark matter halo
  accretion}},
  \href{https://doi.org/10.1103/PhysRevResearch.2.023204}{\emph{Phys. Rev.
  Res.} {\bfseries 2} (2020) 023204}
  [\href{https://arxiv.org/abs/2002.10771}{{\ttfamily 2002.10771}}].

\bibitem{LIGOScientific:2019kan}
{\scshape LIGO Scientific, Virgo} collaboration, B.~P. Abbott et~al.,
  \emph{{Search for Subsolar Mass Ultracompact Binaries in Advanced
  LIGO\textquoteright{}s Second Observing Run}},
  \href{https://doi.org/10.1103/PhysRevLett.123.161102}{\emph{Phys. Rev. Lett.}
  {\bfseries 123} (2019) 161102}
  [\href{https://arxiv.org/abs/1904.08976}{{\ttfamily 1904.08976}}].

\bibitem{Carr:1994ar}
B.~J. Carr, J.~H. Gilbert and J.~E. Lidsey, \emph{{Black hole relics and
  inflation: Limits on blue perturbation spectra}},
  \href{https://doi.org/10.1103/PhysRevD.50.4853}{\emph{Phys. Rev. D}
  {\bfseries 50} (1994) 4853}
  [\href{https://arxiv.org/abs/astro-ph/9405027}{{\ttfamily
  astro-ph/9405027}}].

\bibitem{Schmitz:2020syl}
K.~Schmitz, \emph{{New Sensitivity Curves for Gravitational-Wave Signals from
  Cosmological Phase Transitions}},
  \href{https://doi.org/10.1007/JHEP01(2021)097}{\emph{JHEP} {\bfseries 01}
  (2021) 097} [\href{https://arxiv.org/abs/2002.04615}{{\ttfamily
  2002.04615}}].

\bibitem{Jinno:2017ixd}
R.~Jinno, S.~Lee, H.~Seong and M.~Takimoto, \emph{{Gravitational waves from
  first-order phase transitions: Towards model separation by bubble nucleation
  rate}}, \href{https://doi.org/10.1088/1475-7516/2017/11/050}{\emph{JCAP}
  {\bfseries 11} (2017) 050}
  [\href{https://arxiv.org/abs/1708.01253}{{\ttfamily 1708.01253}}].

\bibitem{Hawking:1971ei}
S.~Hawking, \emph{{Gravitationally collapsed objects of very low mass}},
  \href{https://doi.org/10.1093/mnras/152.1.75}{\emph{Mon. Not. Roy. Astron.
  Soc.} {\bfseries 152} (1971) 75}.

\bibitem{Carr:1974nx}
B.~J. Carr and S.~W. Hawking, \emph{{Black holes in the early Universe}},
  \href{https://doi.org/10.1093/mnras/168.2.399}{\emph{Mon. Not. Roy. Astron.
  Soc.} {\bfseries 168} (1974) 399}.

\bibitem{Ivanov:1994pa}
P.~Ivanov, P.~Naselsky and I.~Novikov, \emph{{Inflation and primordial black
  holes as dark matter}},
  \href{https://doi.org/10.1103/PhysRevD.50.7173}{\emph{Phys. Rev. D}
  {\bfseries 50} (1994) 7173}.

\bibitem{GarciaBellido:1996qt}
J.~Garcia-Bellido, A.~D. Linde and D.~Wands, \emph{{Density perturbations and
  black hole formation in hybrid inflation}},
  \href{https://doi.org/10.1103/PhysRevD.54.6040}{\emph{Phys. Rev. D}
  {\bfseries 54} (1996) 6040}
  [\href{https://arxiv.org/abs/astro-ph/9605094}{{\ttfamily
  astro-ph/9605094}}].

\bibitem{Kawasaki:1997ju}
M.~Kawasaki, N.~Sugiyama and T.~Yanagida, \emph{{Primordial black hole
  formation in a double inflation model in supergravity}},
  \href{https://doi.org/10.1103/PhysRevD.57.6050}{\emph{Phys. Rev. D}
  {\bfseries 57} (1998) 6050}
  [\href{https://arxiv.org/abs/hep-ph/9710259}{{\ttfamily hep-ph/9710259}}].

\bibitem{Yokoyama:1998pt}
J.~Yokoyama, \emph{{Chaotic new inflation and formation of primordial black
  holes}}, \href{https://doi.org/10.1103/PhysRevD.58.083510}{\emph{Phys. Rev.
  D} {\bfseries 58} (1998) 083510}
  [\href{https://arxiv.org/abs/astro-ph/9802357}{{\ttfamily
  astro-ph/9802357}}].

\bibitem{Garcia-Bellido:2017mdw}
J.~Garcia-Bellido and E.~Ruiz~Morales, \emph{{Primordial black holes from
  single field models of inflation}},
  \href{https://doi.org/10.1016/j.dark.2017.09.007}{\emph{Phys. Dark Univ.}
  {\bfseries 18} (2017) 47} [\href{https://arxiv.org/abs/1702.03901}{{\ttfamily
  1702.03901}}].

\bibitem{Hertzberg:2017dkh}
M.~P. Hertzberg and M.~Yamada, \emph{{Primordial Black Holes from Polynomial
  Potentials in Single Field Inflation}},
  \href{https://doi.org/10.1103/PhysRevD.97.083509}{\emph{Phys. Rev. D}
  {\bfseries 97} (2018) 083509}
  [\href{https://arxiv.org/abs/1712.09750}{{\ttfamily 1712.09750}}].

\bibitem{Vilenkin:2000jqa}
A.~Vilenkin and E.~P.~S. Shellard, \emph{{Cosmic Strings and Other Topological
  Defects}}. Cambridge University Press, 7, 2000.

\bibitem{Coleman:1985ki}
S.~R. Coleman, \emph{{Q-balls}},
  \href{https://doi.org/10.1016/0550-3213(86)90520-1}{\emph{Nucl. Phys. B}
  {\bfseries 262} (1985) 263}.

\bibitem{Kusenko:1997si}
A.~Kusenko and M.~E. Shaposhnikov, \emph{{Supersymmetric Q balls as dark
  matter}}, \href{https://doi.org/10.1016/S0370-2693(97)01375-0}{\emph{Phys.
  Lett. B} {\bfseries 418} (1998) 46}
  [\href{https://arxiv.org/abs/hep-ph/9709492}{{\ttfamily hep-ph/9709492}}].

\bibitem{Enqvist:1997si}
K.~Enqvist and J.~McDonald, \emph{{Q balls and baryogenesis in the MSSM}},
  \href{https://doi.org/10.1016/S0370-2693(98)00271-8}{\emph{Phys. Lett. B}
  {\bfseries 425} (1998) 309}
  [\href{https://arxiv.org/abs/hep-ph/9711514}{{\ttfamily hep-ph/9711514}}].

\bibitem{Enqvist:1998en}
K.~Enqvist and J.~McDonald, \emph{{B - ball baryogenesis and the baryon to dark
  matter ratio}},
  \href{https://doi.org/10.1016/S0550-3213(98)00695-6}{\emph{Nucl. Phys. B}
  {\bfseries 538} (1999) 321}
  [\href{https://arxiv.org/abs/hep-ph/9803380}{{\ttfamily hep-ph/9803380}}].

\bibitem{Kasuya:1999wu}
S.~Kasuya and M.~Kawasaki, \emph{{Q ball formation through Affleck-Dine
  mechanism}}, \href{https://doi.org/10.1103/PhysRevD.61.041301}{\emph{Phys.
  Rev. D} {\bfseries 61} (2000) 041301}
  [\href{https://arxiv.org/abs/hep-ph/9909509}{{\ttfamily hep-ph/9909509}}].

\bibitem{Kasuya:2000wx}
S.~Kasuya and M.~Kawasaki, \emph{{Q Ball formation in the gravity mediated SUSY
  breaking scenario}},
  \href{https://doi.org/10.1103/PhysRevD.62.023512}{\emph{Phys. Rev. D}
  {\bfseries 62} (2000) 023512}
  [\href{https://arxiv.org/abs/hep-ph/0002285}{{\ttfamily hep-ph/0002285}}].

\bibitem{Kasuya:2001hg}
S.~Kasuya and M.~Kawasaki, \emph{{Q ball formation: Obstacle to Affleck-Dine
  baryogenesis in the gauge mediated SUSY breaking?}},
  \href{https://doi.org/10.1103/PhysRevD.64.123515}{\emph{Phys. Rev. D}
  {\bfseries 64} (2001) 123515}
  [\href{https://arxiv.org/abs/hep-ph/0106119}{{\ttfamily hep-ph/0106119}}].

\bibitem{Lee:1988ag}
K.-M. Lee, J.~A. Stein-Schabes, R.~Watkins and L.~M. Widrow, \emph{{Gauged q
  Balls}}, \href{https://doi.org/10.1103/PhysRevD.39.1665}{\emph{Phys. Rev. D}
  {\bfseries 39} (1989) 1665}.

\bibitem{Kasuya:2015uka}
S.~Kasuya, M.~Kawasaki and T.~T. Yanagida, \emph{{IceCube potential for
  detecting Q-ball dark matter in gauge mediation}},
  \href{https://doi.org/10.1093/ptep/ptv056}{\emph{PTEP} {\bfseries 2015}
  (2015) 053B02} [\href{https://arxiv.org/abs/1502.00715}{{\ttfamily
  1502.00715}}].

\bibitem{Hong:2015wga}
J.-P. Hong, M.~Kawasaki and M.~Yamada, \emph{{Charged Q-balls in gauge mediated
  SUSY breaking models}},
  \href{https://doi.org/10.1103/PhysRevD.92.063521}{\emph{Phys. Rev. D}
  {\bfseries 92} (2015) 063521}
  [\href{https://arxiv.org/abs/1505.02594}{{\ttfamily 1505.02594}}].

\bibitem{Bogolyubsky:1976nx}
I.~L. Bogolyubsky and V.~G. Makhankov, \emph{{On the Pulsed Soliton Lifetime in
  Two Classical Relativistic Theory Models}}, {\emph{JETP Lett.} {\bfseries 24}
  (1976) 12}.

\bibitem{Segur:1987mg}
H.~Segur and M.~D. Kruskal, \emph{{Nonexistence of Small Amplitude Breather
  Solutions in $\phi^4$ Theory}},
  \href{https://doi.org/10.1103/PhysRevLett.58.747}{\emph{Phys. Rev. Lett.}
  {\bfseries 58} (1987) 747}.

\bibitem{Gleiser:1993pt}
M.~Gleiser, \emph{{Pseudostable bubbles}},
  \href{https://doi.org/10.1103/PhysRevD.49.2978}{\emph{Phys. Rev. D}
  {\bfseries 49} (1994) 2978}
  [\href{https://arxiv.org/abs/hep-ph/9308279}{{\ttfamily hep-ph/9308279}}].

\bibitem{Copeland:1995fq}
E.~J. Copeland, M.~Gleiser and H.~R. Muller, \emph{{Oscillons: Resonant
  configurations during bubble collapse}},
  \href{https://doi.org/10.1103/PhysRevD.52.1920}{\emph{Phys. Rev. D}
  {\bfseries 52} (1995) 1920}
  [\href{https://arxiv.org/abs/hep-ph/9503217}{{\ttfamily hep-ph/9503217}}].

\bibitem{Gleiser:1999tj}
M.~Gleiser and A.~Sornborger, \emph{{Longlived localized field configurations
  in small lattices: Application to oscillons}},
  \href{https://doi.org/10.1103/PhysRevE.62.1368}{\emph{Phys. Rev. E}
  {\bfseries 62} (2000) 1368}
  [\href{https://arxiv.org/abs/patt-sol/9909002}{{\ttfamily
  patt-sol/9909002}}].

\bibitem{Kasuya:2002zs}
S.~Kasuya, M.~Kawasaki and F.~Takahashi, \emph{{I-balls}},
  \href{https://doi.org/10.1016/S0370-2693(03)00344-7}{\emph{Phys. Lett. B}
  {\bfseries 559} (2003) 99}
  [\href{https://arxiv.org/abs/hep-ph/0209358}{{\ttfamily hep-ph/0209358}}].

\bibitem{Gleiser:2004an}
M.~Gleiser, \emph{{d-dimensional oscillating scalar field lumps and the
  dimensionality of space}},
  \href{https://doi.org/10.1016/j.physletb.2004.08.064}{\emph{Phys. Lett. B}
  {\bfseries 600} (2004) 126}
  [\href{https://arxiv.org/abs/hep-th/0408221}{{\ttfamily hep-th/0408221}}].

\bibitem{Fodor:2006zs}
G.~Fodor, P.~Forgacs, P.~Grandclement and I.~Racz, \emph{{Oscillons and
  Quasi-breathers in the phi**4 Klein-Gordon model}},
  \href{https://doi.org/10.1103/PhysRevD.74.124003}{\emph{Phys. Rev. D}
  {\bfseries 74} (2006) 124003}
  [\href{https://arxiv.org/abs/hep-th/0609023}{{\ttfamily hep-th/0609023}}].

\bibitem{Hindmarsh:2006ur}
M.~Hindmarsh and P.~Salmi, \emph{{Numerical investigations of oscillons in 2
  dimensions}}, \href{https://doi.org/10.1103/PhysRevD.74.105005}{\emph{Phys.
  Rev. D} {\bfseries 74} (2006) 105005}
  [\href{https://arxiv.org/abs/hep-th/0606016}{{\ttfamily hep-th/0606016}}].

\bibitem{Amin:2011hj}
M.~A. Amin, R.~Easther, H.~Finkel, R.~Flauger and M.~P. Hertzberg,
  \emph{{Oscillons After Inflation}},
  \href{https://doi.org/10.1103/PhysRevLett.108.241302}{\emph{Phys. Rev. Lett.}
  {\bfseries 108} (2012) 241302}
  [\href{https://arxiv.org/abs/1106.3335}{{\ttfamily 1106.3335}}].

\bibitem{Mukaida:2016hwd}
K.~Mukaida, M.~Takimoto and M.~Yamada, \emph{{On Longevity of
  I-ball/Oscillon}}, \href{https://doi.org/10.1007/JHEP03(2017)122}{\emph{JHEP}
  {\bfseries 03} (2017) 122}
  [\href{https://arxiv.org/abs/1612.07750}{{\ttfamily 1612.07750}}].

\bibitem{Ruffini:1969qy}
R.~Ruffini and S.~Bonazzola, \emph{{Systems of selfgravitating particles in
  general relativity and the concept of an equation of state}},
  \href{https://doi.org/10.1103/PhysRev.187.1767}{\emph{Phys. Rev.} {\bfseries
  187} (1969) 1767}.

\bibitem{Hogan:1988mp}
C.~J. Hogan and M.~J. Rees, \emph{{AXION MINICLUSTERS}},
  \href{https://doi.org/10.1016/0370-2693(88)91655-3}{\emph{Phys. Lett. B}
  {\bfseries 205} (1988) 228}.

\bibitem{Kolb:1993zz}
E.~W. Kolb and I.~I. Tkachev, \emph{{Axion miniclusters and Bose stars}},
  \href{https://doi.org/10.1103/PhysRevLett.71.3051}{\emph{Phys. Rev. Lett.}
  {\bfseries 71} (1993) 3051}
  [\href{https://arxiv.org/abs/hep-ph/9303313}{{\ttfamily hep-ph/9303313}}].

\bibitem{Seidel:1993zk}
E.~Seidel and W.-M. Suen, \emph{{Formation of solitonic stars through
  gravitational cooling}},
  \href{https://doi.org/10.1103/PhysRevLett.72.2516}{\emph{Phys. Rev. Lett.}
  {\bfseries 72} (1994) 2516}
  [\href{https://arxiv.org/abs/gr-qc/9309015}{{\ttfamily gr-qc/9309015}}].

\bibitem{Hu:2000ke}
W.~Hu, R.~Barkana and A.~Gruzinov, \emph{{Cold and fuzzy dark matter}},
  \href{https://doi.org/10.1103/PhysRevLett.85.1158}{\emph{Phys. Rev. Lett.}
  {\bfseries 85} (2000) 1158}
  [\href{https://arxiv.org/abs/astro-ph/0003365}{{\ttfamily
  astro-ph/0003365}}].

\bibitem{Guzman:2006yc}
F.~S. Guzman and L.~A. Urena-Lopez, \emph{{Gravitational cooling of
  self-gravitating Bose-Condensates}},
  \href{https://doi.org/10.1086/504508}{\emph{Astrophys. J.} {\bfseries 645}
  (2006) 814} [\href{https://arxiv.org/abs/astro-ph/0603613}{{\ttfamily
  astro-ph/0603613}}].

\bibitem{Sikivie:2009qn}
P.~Sikivie and Q.~Yang, \emph{{Bose-Einstein Condensation of Dark Matter
  Axions}}, \href{https://doi.org/10.1103/PhysRevLett.103.111301}{\emph{Phys.
  Rev. Lett.} {\bfseries 103} (2009) 111301}
  [\href{https://arxiv.org/abs/0901.1106}{{\ttfamily 0901.1106}}].

\bibitem{Liebling:2012fv}
S.~L. Liebling and C.~Palenzuela, \emph{{Dynamical boson stars}},
  \href{https://doi.org/10.1007/s41114-023-00043-4}{\emph{Living Rev. Rel.}
  {\bfseries 26} (2023) 1} [\href{https://arxiv.org/abs/1202.5809}{{\ttfamily
  1202.5809}}].

\bibitem{Guth:2014hsa}
A.~H. Guth, M.~P. Hertzberg and C.~Prescod-Weinstein, \emph{{Do Dark Matter
  Axions Form a Condensate with Long-Range Correlation?}},
  \href{https://doi.org/10.1103/PhysRevD.92.103513}{\emph{Phys. Rev. D}
  {\bfseries 92} (2015) 103513}
  [\href{https://arxiv.org/abs/1412.5930}{{\ttfamily 1412.5930}}].

\bibitem{Eby:2014fya}
J.~Eby, P.~Suranyi, C.~Vaz and L.~C.~R. Wijewardhana, \emph{{Axion Stars in the
  Infrared Limit}}, \href{https://doi.org/10.1007/JHEP11(2016)134}{\emph{JHEP}
  {\bfseries 03} (2015) 080} [\href{https://arxiv.org/abs/1412.3430}{{\ttfamily
  1412.3430}}].

\bibitem{Braaten:2015eeu}
E.~Braaten, A.~Mohapatra and H.~Zhang, \emph{{Dense Axion Stars}},
  \href{https://doi.org/10.1103/PhysRevLett.117.121801}{\emph{Phys. Rev. Lett.}
  {\bfseries 117} (2016) 121801}
  [\href{https://arxiv.org/abs/1512.00108}{{\ttfamily 1512.00108}}].

\bibitem{Braaten:2016kzc}
E.~Braaten, A.~Mohapatra and H.~Zhang, \emph{{Nonrelativistic Effective Field
  Theory for Axions}},
  \href{https://doi.org/10.1103/PhysRevD.94.076004}{\emph{Phys. Rev. D}
  {\bfseries 94} (2016) 076004}
  [\href{https://arxiv.org/abs/1604.00669}{{\ttfamily 1604.00669}}].

\bibitem{Hui:2016ltb}
L.~Hui, J.~P. Ostriker, S.~Tremaine and E.~Witten, \emph{{Ultralight scalars as
  cosmological dark matter}},
  \href{https://doi.org/10.1103/PhysRevD.95.043541}{\emph{Phys. Rev. D}
  {\bfseries 95} (2017) 043541}
  [\href{https://arxiv.org/abs/1610.08297}{{\ttfamily 1610.08297}}].

\bibitem{Eby:2016cnq}
J.~Eby, M.~Leembruggen, P.~Suranyi and L.~C.~R. Wijewardhana, \emph{{Collapse
  of Axion Stars}}, \href{https://doi.org/10.1007/JHEP12(2016)066}{\emph{JHEP}
  {\bfseries 12} (2016) 066}
  [\href{https://arxiv.org/abs/1608.06911}{{\ttfamily 1608.06911}}].

\bibitem{Eby:2017teq}
J.~Eby, P.~Suranyi and L.~C.~R. Wijewardhana, \emph{{Expansion in Higher
  Harmonics of Boson Stars using a Generalized Ruffini-Bonazzola Approach, Part
  1: Bound States}},
  \href{https://doi.org/10.1088/1475-7516/2018/04/038}{\emph{JCAP} {\bfseries
  04} (2018) 038} [\href{https://arxiv.org/abs/1712.04941}{{\ttfamily
  1712.04941}}].

\bibitem{Eby:2018ufi}
J.~Eby, K.~Mukaida, M.~Takimoto, L.~C.~R. Wijewardhana and M.~Yamada,
  \emph{{Classical nonrelativistic effective field theory and the role of
  gravitational interactions}},
  \href{https://doi.org/10.1103/PhysRevD.99.123503}{\emph{Phys. Rev. D}
  {\bfseries 99} (2019) 123503}
  [\href{https://arxiv.org/abs/1807.09795}{{\ttfamily 1807.09795}}].

\bibitem{Mazur:2001fv}
P.~O. Mazur and E.~Mottola, \emph{{Gravitational Condensate Stars: An
  Alternative to Black Holes}},
  \href{https://doi.org/10.3390/universe9020088}{\emph{Universe} {\bfseries 9}
  (2023) 88} [\href{https://arxiv.org/abs/gr-qc/0109035}{{\ttfamily
  gr-qc/0109035}}].

\bibitem{Mazur:2004fk}
P.~O. Mazur and E.~Mottola, \emph{{Gravitational vacuum condensate stars}},
  \href{https://doi.org/10.1073/pnas.0402717101}{\emph{Proc. Nat. Acad. Sci.}
  {\bfseries 101} (2004) 9545}
  [\href{https://arxiv.org/abs/gr-qc/0407075}{{\ttfamily gr-qc/0407075}}].

\bibitem{Carballo-Rubio:2017tlh}
R.~Carballo-Rubio, \emph{{Stellar equilibrium in semiclassical gravity}},
  \href{https://doi.org/10.1103/PhysRevLett.120.061102}{\emph{Phys. Rev. Lett.}
  {\bfseries 120} (2018) 061102}
  [\href{https://arxiv.org/abs/1706.05379}{{\ttfamily 1706.05379}}].

\bibitem{Aharonov:1987tp}
Y.~Aharonov, A.~Casher and S.~Nussinov, \emph{{The Unitarity Puzzle and Planck
  Mass Stable Particles}},
  \href{https://doi.org/10.1016/0370-2693(87)91320-7}{\emph{Phys. Lett. B}
  {\bfseries 191} (1987) 51}.

\bibitem{Adler:2001vs}
R.~J. Adler, P.~Chen and D.~I. Santiago, \emph{{The Generalized uncertainty
  principle and black hole remnants}},
  \href{https://doi.org/10.1023/A:1015281430411}{\emph{Gen. Rel. Grav.}
  {\bfseries 33} (2001) 2101}
  [\href{https://arxiv.org/abs/gr-qc/0106080}{{\ttfamily gr-qc/0106080}}].

\bibitem{Balkin:2021zfd}
R.~Balkin, J.~Serra, K.~Springmann, S.~Stelzl and A.~Weiler, \emph{{Density
  induced vacuum instability}},
  \href{https://doi.org/10.21468/SciPostPhys.14.4.071}{\emph{SciPost Phys.}
  {\bfseries 14} (2023) 071}
  [\href{https://arxiv.org/abs/2105.13354}{{\ttfamily 2105.13354}}].

\bibitem{Kusenko:1997hj}
A.~Kusenko, \emph{{Phase transitions precipitated by solitosynthesis}},
  \href{https://doi.org/10.1016/S0370-2693(97)00700-4}{\emph{Phys. Lett. B}
  {\bfseries 406} (1997) 26}
  [\href{https://arxiv.org/abs/hep-ph/9705361}{{\ttfamily hep-ph/9705361}}].

\bibitem{Metaxas:2000qf}
D.~Metaxas, \emph{{Nontopological solitons as nucleation sites for cosmological
  phase transitions}},
  \href{https://doi.org/10.1103/PhysRevD.63.083507}{\emph{Phys. Rev. D}
  {\bfseries 63} (2001) 083507}
  [\href{https://arxiv.org/abs/hep-ph/0009225}{{\ttfamily hep-ph/0009225}}].

\bibitem{Pearce:2012jp}
L.~Pearce, \emph{{Solitosynthesis induced phase transitions}},
  \href{https://doi.org/10.1103/PhysRevD.85.125022}{\emph{Phys. Rev. D}
  {\bfseries 85} (2012) 125022}
  [\href{https://arxiv.org/abs/1202.0873}{{\ttfamily 1202.0873}}].

\bibitem{Steinhardt:1981ec}
P.~J. Steinhardt, \emph{{Monopole and Vortex Dissociation and Decay of the
  False Vacuum}},
  \href{https://doi.org/10.1016/0550-3213(81)90449-1}{\emph{Nucl. Phys. B}
  {\bfseries 190} (1981) 583}.

\bibitem{Steinhardt:1981mm}
P.~J. Steinhardt, \emph{{Monopole Dissociation in the Early Universe}},
  \href{https://doi.org/10.1103/PhysRevD.24.842}{\emph{Phys. Rev. D} {\bfseries
  24} (1981) 842}.

\bibitem{Jensen:1982jv}
L.~G. Jensen and P.~J. Steinhardt, \emph{{DISSOCIATION OF
  ABRIKOSOV-NIELSEN-OLESEN VORTICES}},
  \href{https://doi.org/10.1103/PhysRevB.27.5549}{\emph{Phys. Rev. B}
  {\bfseries 27} (1983) 5549}.

\bibitem{Witten:1984rs}
E.~Witten, \emph{{Cosmic Separation of Phases}},
  \href{https://doi.org/10.1103/PhysRevD.30.272}{\emph{Phys. Rev. D} {\bfseries
  30} (1984) 272}.

\bibitem{Yajnik:1986tg}
U.~A. Yajnik, \emph{{PHASE TRANSITION INDUCED BY COSMIC STRINGS}},
  \href{https://doi.org/10.1103/PhysRevD.34.1237}{\emph{Phys. Rev. D}
  {\bfseries 34} (1986) 1237}.

\bibitem{Yajnik:1986wq}
U.~A. Yajnik and T.~Padmanabhan, \emph{{ANALYTICAL APPROACH TO STRING INDUCED
  PHASE TRANSITION}},
  \href{https://doi.org/10.1103/PhysRevD.35.3100}{\emph{Phys. Rev. D}
  {\bfseries 35} (1987) 3100}.

\bibitem{Dasgupta:1997kn}
I.~Dasgupta, \emph{{Vacuum tunneling by cosmic strings}},
  \href{https://doi.org/10.1016/S0550-3213(97)00546-4}{\emph{Nucl. Phys. B}
  {\bfseries 506} (1997) 421}
  [\href{https://arxiv.org/abs/hep-th/9702041}{{\ttfamily hep-th/9702041}}].

\bibitem{Kumar:2008jb}
B.~Kumar and U.~A. Yajnik, \emph{{On stability of false vacuum in
  supersymmetric theories with cosmic strings}},
  \href{https://doi.org/10.1103/PhysRevD.79.065001}{\emph{Phys. Rev. D}
  {\bfseries 79} (2009) 065001}
  [\href{https://arxiv.org/abs/0807.3254}{{\ttfamily 0807.3254}}].

\bibitem{Kumar:2009pr}
B.~Kumar and U.~Yajnik, \emph{{Graceful exit via monopoles in a theory with
  O'Raifeartaigh type supersymmetry breaking}},
  \href{https://doi.org/10.1016/j.nuclphysb.2010.01.011}{\emph{Nucl. Phys. B}
  {\bfseries 831} (2010) 162}
  [\href{https://arxiv.org/abs/0908.3949}{{\ttfamily 0908.3949}}].

\bibitem{Kumar:2010mv}
B.~Kumar, M.~B. Paranjape and U.~A. Yajnik, \emph{{Fate of the false monopoles:
  Induced vacuum decay}},
  \href{https://doi.org/10.1103/PhysRevD.82.025022}{\emph{Phys. Rev. D}
  {\bfseries 82} (2010) 025022}
  [\href{https://arxiv.org/abs/1006.0693}{{\ttfamily 1006.0693}}].

\bibitem{Lee:2013zca}
B.-H. Lee, W.~Lee, R.~MacKenzie, M.~B. Paranjape, U.~A. Yajnik and D.-h. Yeom,
  \emph{{Battle of the bulge: Decay of the thin, false cosmic string}},
  \href{https://doi.org/10.1103/PhysRevD.88.105008}{\emph{Phys. Rev. D}
  {\bfseries 88} (2013) 105008}
  [\href{https://arxiv.org/abs/1310.3005}{{\ttfamily 1310.3005}}].

\bibitem{Lee:2013ega}
B.-H. Lee, W.~Lee, R.~MacKenzie, M.~B. Paranjape, U.~A. Yajnik and D.-h. Yeom,
  \emph{{Tunneling decay of false vortices}},
  \href{https://doi.org/10.1103/PhysRevD.88.085031}{\emph{Phys. Rev. D}
  {\bfseries 88} (2013) 085031}
  [\href{https://arxiv.org/abs/1308.3501}{{\ttfamily 1308.3501}}].

\bibitem{Koga:2019mee}
I.~Koga, S.~Kuroyanagi and Y.~Ookouchi, \emph{{Instability of Higgs Vacuum via
  String Cloud}},
  \href{https://doi.org/10.1016/j.physletb.2019.135093}{\emph{Phys. Lett. B}
  {\bfseries 800} (2020) 135093}
  [\href{https://arxiv.org/abs/1910.02435}{{\ttfamily 1910.02435}}].

\bibitem{Agrawal:2022hnf}
P.~Agrawal and M.~Nee, \emph{{The Boring Monopole}},
  \href{https://doi.org/10.21468/SciPostPhys.13.3.049}{\emph{SciPost Phys.}
  {\bfseries 13} (2022) 049}
  [\href{https://arxiv.org/abs/2202.11102}{{\ttfamily 2202.11102}}].

\bibitem{Blasi:2022woz}
S.~Blasi and A.~Mariotti, \emph{{Domain Walls Seeding the Electroweak Phase
  Transition}},
  \href{https://doi.org/10.1103/PhysRevLett.129.261303}{\emph{Phys. Rev. Lett.}
  {\bfseries 129} (2022) 261303}
  [\href{https://arxiv.org/abs/2203.16450}{{\ttfamily 2203.16450}}].

\bibitem{Blasi:2023rqi}
S.~Blasi, R.~Jinno, T.~Konstandin, H.~Rubira and I.~Stomberg,
  \emph{{Gravitational waves from defect-driven phase transitions: domain
  walls}},  \href{https://arxiv.org/abs/2302.06952}{{\ttfamily 2302.06952}}.

\end{thebibliography}\endgroup


\end{document}